\documentclass{article}
\usepackage[preprint]{tmlr}
\usepackage[tmlr,subfig,algpseudocodex]{definition}

\PassOptionsToPackage{numbers, compress}{natbib}
\usepackage[version=4]{mhchem}
\usepackage[toc,page,header]{appendix}

\newcommand{\ours}{\textsf{MatLLMSearch}\xspace}
\title{\ours: Crystal Structure Discovery with Evolution-Guided Large Language Models}

\author{\name Jingru Gan \email jrgan@cs.ucla.edu \\
      \addr University of California, Los Angeles
      \AND
      \name Peichen Zhong \\
      \addr University of California, Berkeley
      \AND
      \name Yuanqi Du \\
      \addr Cornell University
      \AND
      \name Yanqiao Zhu \\
      \addr University of California, Los Angeles
      \AND
      \name Chenru Duan \\
      \addr Deep Principle Inc.
      \AND
      \name Haorui Wang \\
      \addr Georgia Institute of Technology
      \AND
      \name Daniel Schwalbe-Koda \\
      \addr University of California, Los Angeles
      \AND
      \name Carla P. Gomes \\
      \addr Cornell University
      \AND
      \name Kristin A. Persson \\
      \addr University of California, Berkeley
      \AND
      \name Wei Wang \\
      \addr University of California, Los Angeles
}

\begin{document}

\maketitle

\begin{abstract}

Crystal structure generation is fundamental to materials science, enabling the discovery of novel materials with desired properties. While existing approaches leverage Large Language Models (LLMs) through extensive fine-tuning on materials databases, we show that pre-trained LLMs can inherently generate novel and stable crystal structures without additional fine-tuning.
Our framework employs LLMs as intelligent proposal agents within an evolutionary pipeline that guides them to perform implicit crossover and mutation operations while maintaining chemical validity. We demonstrate that \ours achieves a 78.38\% metastable rate validated by machine learning interatomic potentials and 31.7\% DFT-verified stability, outperforming specialized models such as CrystalTextLLM.
Beyond crystal structure generation, we further demonstrate that our framework adapts to diverse materials design tasks, including crystal structure prediction and multi-objective optimization of properties such as deformation energy and bulk modulus, all without fine-tuning.
These results establish our framework as a versatile and effective framework for consistent high-quality materials discovery, offering training-free generation of novel stable structures with reduced overhead and broader accessibility.

\end{abstract}

\section{Introduction}

Crystal Structure Generation (CSG) and Prediction (CSP) represent critical bottlenecks in materials discovery, requiring both chemical validity and thermodynamic stability to determine whether a material can be synthesized \cite{bagayoko2014dft}. These tasks demand navigating an expansive chemical space while satisfying multiple constraints: three-dimensional periodicity, proper atomic coordination, charge balance, and minimized formation energy. While computational approaches have emerged as indispensable tools for accelerating materials discovery \cite{dunn2020benchmarking, eremin2023gnn}, developing reliable systems that effectively explore this vast and complex space remains challenging.

Recent advances in deep learning have introduced various approaches for structure prediction, from variational autoencoders to diffusion models \cite{xie2022crystal,Zeni2025MatterGen,flam2023language,jiao2024crystal,gruver2024llmtime}. Meanwhile, Large Language Models (LLMs) have emerged as powerful tools for materials discovery \cite{achiam2023gpt,Antunes2023CrystalSG,Fu_2023}. Prior work \cite{flam2023language} demonstrated that autoregressive models using character-level tokenization can generate valid crystal structures, and \citet{gruver2024llmtime} showing that fine-tuning pre-trained language models like Llama \cite{grattafiori2024llama3herdmodels} can produce physically stable structures.

Current approaches often fine-tune LLMs on materials databases such as the Materials Project \cite{gruver2024llmtime}, which contains only tens of thousands of structures compared to the vast space of possible stable compounds. While we propose a fundamentally different perspective: Recognizing pre-trained LLMs not as tools requiring domain-specific fine-tuning, but as intelligent agents already possessing rich embedded knowledge from vast scientific corpora. This perspective raises the question: \textit{How can we exploit the comprehensive scientific knowledge already embedded in pre-trained LLMs to build a system that can consistently produce valid stable crystal structures?} 

Intuitively, we may directly prompt a commercial LLM to generate crystal structures. However, our ablational experiments in \cref{sec:exp_ablation} and \cref{appendix:ablation} across multiple configurations reveal that simple prompting fails to consistently generate valid crystal structures that are both stable and novel. These attempts often produce either copies of known structures, chemically invalid configurations, or thermodynamically unstable structures. The failure suggests that LLMs struggle to simultaneously satisfy the multiple constraints of CSG, indicating the need for a more sophisticated approach to exploit the potential of LLMs for materials discovery.

Evolutionary algorithms provide an effective framework for exploring the vast chemical space \cite{oganov2006evolutionary, allahyari2020coevolutionary, wang2024efficientevolutionarysearchchemical}. By mimicking biological evolution through iterative selection, reproduction, and mutation operations, these algorithms can gradually improve the candidates, enabling automated property-guided materials optimization. Previous evolutionary approaches to CSG and CSP rely on explicit crossover and mutation operators, such as swapping structural motifs or introducing atomic displacements. While effective, these traditional operators lack the chemical intuition to efficiently navigate the complex constraints of crystal structures, often resulting in physically implausible candidates.

Our work advances this paradigm by leveraging the rich scientific knowledge embedded in LLMs to perform chemically-informed operations within the evolutionary algorithm framework. 
Unlike traditional operators that manipulate structures based on predefined rules, LLMs can implicitly reason about chemical bonding patterns, structural motifs, and stability principles learned from vast scientific literature \cite{Bran:2024dw,Guo:2023wn,Boiko:2023ky}. This knowledge-guided approach enables more intelligent exploration of the chemical space, potentially discovering novel structures that traditional evolutionary methods might miss due to their limited chemical knowledge.

\begin{figure*}[t]
    \centering
    \vskip-1em
    \includegraphics[width=\textwidth]{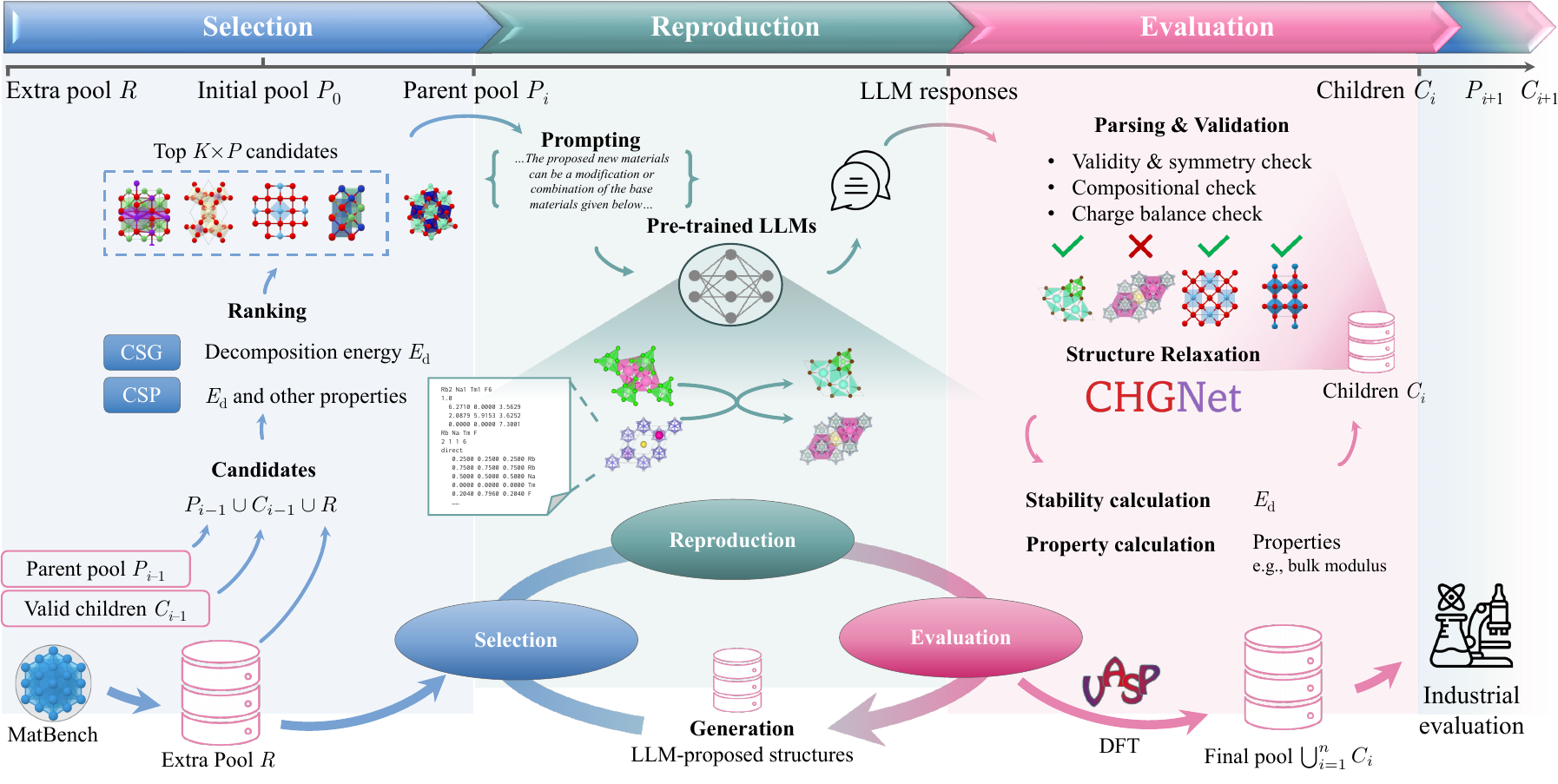}
    \caption{The workflow of \ours for crystal structure generation. Starting from an initial population of known structures, our framework iteratively evolves new crystal structures through LLM-guided reproduction, evaluation, and selection.}
    \label{fig:pipeline}
\end{figure*}

In this work, we introduce \ours, a novel framework that integrates the rich scientific knowledge of LLMs into the evolutionary framework for crystal structure discovery. In our proposed framework, LLMs function as intelligent proposal agents that analyze parent structures to perform implicit crossover and mutation operations, while Machine Learning Interatomic Potentials (MLIPs) evaluate the physicochemical validity of generated structures.
As illustrated in \cref{fig:pipeline}, through iterative selection, reproduction, and evaluation, \ours progressively discovers crystal structures with desired properties.

Our experiments demonstrate that \ours successfully generates diverse, thermodynamically stable crystal structures while maintaining crystallographic validity. The framework achieves a 76.8\% metastable structure generation rate, with 31.7\% of structures verified as stable through DFT calculations, surpassing the state-of-the-art fine-tuned model CrystalTextLLM \cite{gruver2024llmtime}. Notably, this performance is achieved with minimal computational overhead, requiring only LLM inference and stability evaluation with MLIPs rather than extensive model training.

Beyond crystal structure generation, our framework demonstrates remarkable flexibility across various materials discovery tasks. Through simple modifications in prompting and reference seed structures selection, our method extends to CSP, which we validate by discovering several metastable \ce{Na3AlCl6} polymorphs with significantly higher stability than existing structures in the Materials Project database. Furthermore, the framework enables multi-objective optimization of properties such as bulk modulus, without requiring specialized fine-tuning. While we demonstrate results using general-purpose pre-trained LLMs, the framework could also incorporate domain-specialized fine-tuned models or alternative search algorithms, offering a computationally efficient approach to materials discovery with reduced overhead and broader accessibility.

\section{Background: Computational Materials Discovery with Machine Learning}

\subsection{Problem Definition}
\label{sec:task}

\textbf{Crystal Structure Generation (CSG).} 
The objective of CSG is to learn a probability distribution $p(c, l, s)$ over crystalline materials, where $c \in \mathbb{R}^{N \times K}$ represents the chemical composition matrix for $N$ atoms of $K$ distinct chemical species, $l \in \mathbb{R}^{6}$ denotes the lattice parameters (lengths and angles), and $s \in \mathbb{R}^{N \times 3}$ defines the spatial coordinates of atoms within a periodic unit cell.
Samples drawn from this distribution should ideally satisfy fundamental thermodynamic stability criteria. %

\textbf{Crystal Structure Prediction (CSP).} CSP addresses a more constrained problem of determining stable crystal structures for a specified chemical composition. Formally, it learns a conditional probability distribution $p(s, l\mid c)$ to identify thermodynamically favorable atomic arrangements and lattice parameters given a fixed composition $c$. This formulation addresses the practical scenario of discovering stable polymorphs for a specified chemical formula.

\textbf{Crystal Structure Design (CSD).} CSD extends beyond structure prediction by incorporating property optimization and conditional generation. An example objective is finding the optimal crystal structure that maximizes a target property $h(c, l, s)$: $m^* = \argmax_{c, l, s \sim p(c, l, s)} h(c, l, s)$, where $h:\mathbb{R}^{N\times K} \times \mathbb{R}^{6} \times \mathbb{R}^{N\times 3} \rightarrow \mathbb{R}$ represents an oracle function evaluating the desired materials property. It can also be formulated as sampling from a tilted distribution $p(c,l, s)\,\exp(h(c, l, s))$ \cite{rafailov2024direct}. 
Additional constraints can be integrated into the design process, allowing for flexible tasks such as compositional substitution (learning $p(c\mid l, s)$) and composition/structure completion (inpainting generation, learning $p(c^{\text{unknown}}, s^{\text{unknown}}\mid c^{\text{known}}, l,  s^{\text{known}})$)  \cite{dai2024inpainting}.

\subsection{(Meta)Stability of Materials}
\label{sec:stability}
Among computational approaches for evaluating crystal structure stability, Density Functional Theory (DFT) is the most reliable method for predicting formation energies in solid-state materials, showing close alignment with experimental measurements \cite{Jain2011_GGAU, sun2016dft}.
The thermodynamic stability of a structure is quantified through its decomposition energy ($E_\text{d}$) with respect to the convex hull of known stable phases: $E_\text{d} = E_{\mathrm{s}} - \sum_i x_i E_i$, where $E_{\mathrm{s}}$ represents the total energy per atom, $x_i$ denotes the molar fraction of the $i$-th competing phase, and $E_i$ corresponds to its ground-state energy per atom. While the convex hull serves as a fixed reference, the evaluated structure $s$ need not be part of this hull. A negative decomposition energy ($E_\text{d} < 0$) indicates a thermodynamically stable state below the convex hull, while $E_\text{d} > 0$ suggests a metastable phase with a driving force for decomposition into more stable compounds. Our main objective for CSG is to identify stable crystal structures where $E_\text{d} \leq 0$.

Given the computational intensity of DFT calculations, universal Machine Learning Interatomic Potentials (MLIPs), trained on millions of DFT calculations, have emerged as efficient and reliable proxies for structure stability assessment. 
Notable among these is CHGNet\cite{deng2023chgnet}, a Graph Neural Network (GNN)-based MLIP that uniquely incorporates magnetic moments to capture both atomic and electronic interactions. M3GNet\cite{chen2022m3gnet} offers an alternative approach, implementing three-body interactions in its graph architecture for accurate structural predictions across diverse chemical spaces.
Recent advances in universal MLIPs include MACE \cite{batatia2023foundation}, DPA-1 \cite{zhang2024dpa1}, and JMP \cite{shoghi2023molecules}, which demonstrate high accuracy in predicting crystal thermodynamic stability, particularly when trained on industrial-scale datasets comprising millions of compounds and non-equilibrium atomic configurations \cite{merchant2023GNoME, barroso2024open, yang2024mattersim}.
In this work, we employ the pre-trained CHGNet as our universal MLIP due to its closer alignment with DFT results, using a fixed phase diagram derived from the Materials Project 2023 DFT calculations \cite{wang2021dft, Jain2011_GGAU}.

\section{\ours}
\label{sec:method}

We propose \ours, an evolutionary workflow that leverages pre-trained LLMs to search for stable and optimized crystal structures with. In this section, we introduce three key stages of the workflow as illustrated in \cref{fig:pipeline}: (1) \textbf{Selection}, which identifies promising candidate structures from existing pools based on stability and property metrics; (2) \textbf{Reproduction}, where the LLM generates new candidates through implicit crossover and mutations of parent structures; and (3) \textbf{Evaluation}, which assesses proposed structures for validity, stability, and target properties. The overall workflow, outlined in \cref{appendix:algorithm}, iteratively evolves a population of crystal structures while maintaining physical constraints and optimizing desired properties.

\subsection{Initialization}
Our evolutionary search begins by sampling $(K \times P)$ structures from a database of known stable structures $\mathcal{D}$ to form our initial parent pool $\mathcal{P}_{0}$. These structures are organized into $K$ groups of $P$ parents each to serve as reference examples in LLM prompts. We optionally retrieve an extra pool $\mathcal{R}$ from $\mathcal{D}$ to expand the candidate space during selection. $\mathcal{R}$ can be customized to suit various design objectives, with more details and ablation studies provided in \cref{sec:exp_ablation}.

\subsection{Reproduction} 
Genetic algorithms traditionally mimic biological evolution through explicit crossover and mutation operations \cite{johnston2003evolving,heiles2013global}. In crystal structure prediction, crossover typically involves combining structural fragments from parent structures (e.g., swapping atomic positions or structural motifs), while mutation introduces random variations through predefined operations like atomic displacement, lattice transformation, or element substitution \cite{kadan2023GAmuza,curtis2018GAtor}. While effective, these rigid operators can limit the exploration of the complex crystal structure space.
In \ours, we explore the flexibility of LLMs for structure reproduction. Through prompt-based guidance, we ask LLMs to perform implicit crossover and mutation by analyzing and combining structural information from parent materials. Specifically, LLMs are instructed to ``modify or combine the base materials'', while maintaining chemical validity and enhancing target properties.
This approach allows LLMs to freely and simultaneously introduce variations across multiple structural aspects, including atomic positions, lattice parameters, and element substitutions, or even generate completely new structures functionally relevant to parent structures.

\subsection{Evaluation}
\label{sec:method_eval}
Our evaluation pipeline consists of two stages:
\begin{itemize}
    \item \textbf{Rule-based validation} ensures structural integrity by verifying three-dimensional periodicity, physical connectivity (interatomic distances between 0.6-1.3 times the sum of atomic radii), and chemical validity through charge balance analysis.
    \item \textbf{Stability and property evaluation} begins with structure relaxation using CHGNet. We quantify thermodynamic stability through decomposition energy $E_\text{d}$ calculated as the distance to the Materials Project convex hull. Notably, we observe that LLM-proposed structures typically require minimal relaxation, with 61.1\% of structures exhibiting small energy changes ($|\Delta E| < 0.5$ eV/atom) during this process (detailed in \cref{appendix:relaxation}), indicating their initial stability.
For stability-focused optimization, we quantify thermodynamic stability through the decomposition energy $E_\text{d}$ using CHGNet, calculated as the distance to the convex hull from the Materials Project database (version \verb|2023-02-07-ppd-mp|).
For multi-objective optimization, additional properties such as bulk modulus can be evaluated. 
These quantitative scores then guide the selection process for subsequent generations, allowing our framework to flexibly adapt to different design goals.
\end{itemize}

\subsection{Selection}
Last, the selection stage evolves a population of candidate structures that meet the optimization objectives, such as thermodynamic stability or other desired physical properties.
For each iteration $i$, we construct a new parent pool $\mathcal{P}_{i+1}$ of the same size $(K \times P)$ by selecting top-ranked candidates from three sources: the current parent pool ($\mathcal{P}_i$), newly generated children structures ($\mathcal{C}_{i}$), and an optional extra pool ($\mathcal{R}$) to improve diversity.
Candidates in $\mathcal{P}_{i} \cup \mathcal{C}_{i} \cup \mathcal{R}$ are ranked according to optimization objectives, either single-objective (e.g., $E_\text{d}$ for stability) or multi-objective criteria (e.g., alternating between different properties).

\subsection{Final DFT Verification}

After completing all evolutionary iterations, we collect the cumulated offspring structures $\mathcal{S} = \bigcup_i \mathcal{C}_i$ for final validation using DFT.
To save computational cost, we focus on meta-stable structures with CHGNet-predicted decomposition energy $E_d < 0.1$~eV/atom.
DFT calculations are performed using VASP 6 in the Generalized Gradient Approximation (GGA) with PBE functional \cite{perdew1996_GGA}, using the projector-augmented wave method \cite{Kresse1996_VASP, Kresse1999_PAW}. We employed a plane-wave basis set with an energy cutoff of 520 eV and a $k$-point mesh of 1,000 per reciprocal atom \cite{Jain2013MP}. The calculations converged to $10^{-6}$ eV in total energy for electronic self-consistent field cycles and 0.02 eV/Å in interatomic forces for the ionic steps. The computational settings are consistent with MPGGARelaxSet and MPGGAStaticSet \cite{Jain2011_GGAU}.

\section{Experiments}
\subsection{Experimental Settings}
\label{sec:exp_settings}
We use Llama 3.1 (70B)~\cite{grattafiori2024llama3herdmodels} as the base LLM. We set temperature to 0.95 to balance creativity and reliability. All experiments use parent size $P=2$, reproduction size $C=5$, and $N=10$ iterations, with population size $K=100$ unless otherwise specified. Crystal structures are represented in POSCAR format with 12 decimal digits.

For initialization, we use the MatBench dataset~\cite{dunn2020benchmarking} as the known stable structure set $\mathcal{D}$. 
We sample some known stable structures from $\mathcal{D}$ to form the initial generation and fill the parents during each iteration. These samples are chosen based on their CHGNet-predicted decomposition energy or the properties to be optimized. 
Detailed ablation studies regarding this selection policy are provided in \cref{appendix:exp_details}.

\begin{table*}[t]
    \vskip-1em
    \centering
    \small
    \SetTblrInner{rowsep=1.3pt}
    \begin{tblr}{
    colspec = {ccccccccc},
        row{1-3} = {bg=gray!25},
        cell{odd[4-22]}{2-9} = {bg=gray!10}
        }
        \toprule
        \SetCell[r=3]{c}{\textbf{Model}} & 
        \SetCell[r=3]{c}{\textbf{$f$-ele in}\\ \textbf{Parents$^\dagger$}} & 
        \SetCell[c=2]{c}{\textbf{Validity}} & & 
        \SetCell[c=3]{c}{\textbf{Metastability}} & & &
        \SetCell[c=2]{c}{\textbf{Stability$^\ddagger$}} &\\
        \cmidrule[lr]{3-4} \cmidrule[lr]{5-7} \cmidrule[lr]{8-9} 
        & & \SetCell[r=2]{c}{Structural} & \SetCell[r=2]{c}{Composition} & M3GNet & \SetCell[c=2]{c}{CHGNet} & & \SetCell[c=2]{c}{DFT} &\\
        \cmidrule[lr]{5} \cmidrule[lr]{6-7} \cmidrule[lr]{8-9} 
        & & & & $E_{\text{d}} < 0.1$ & $E_{\text{d}}<0.1$ & $E_{\text{d}}<0.03$ &  w/ $f$-ele & w/o $f$-ele$^\S$ \\
        \midrule
        CDVAE$^*$ & ---  & 100.0\% & 86.7\% & 28.8\% & --- & --- & 5.4\% & --- \\
        \hline[dashed]
        CrystalTextLLM-7B$^*$ & --- & 96.4\% & 93.3\% & 35.0\% & --- & --- & 8.7\% & --- \\
        CrystalTextLLM-13B$^*$ & --- & 95.5\% & 92.4\% & 38.0\% & --- & --- & 14.4\% & --- \\
        CrystalTextLLM-70B$^*$ & --- & 99.6\% & 95.4\% & 49.8\% & --- & --- & 10.6\% & --- \\
        \hline[dashed]
        \SetCell[r=2]{c}{\ours \\ (Llama 3.1-70B)} & \cmark & 100.0\% & 79.4\% & 81.1\% & 76.8\% & 56.5\% & 31.7\% & 14.0\% \\
        & \xmark & 100.0\% & 89.0\% & 81.9\% & {78.4\%} & 54.8\% & 27.0\% & 24.6\% \\
        \bottomrule
        \end{tblr}
        \caption{Performance comparison of crystal structure generation. Metastability is first assessed using surrogate models, where we report both M3GNet and CHGNet results for fair comparison with baselines CDVAE and CrystalTextLLM (which use M3GNet). $^*$Results taken from the original papers. $^\dagger$Indicates whether $f$-electron elements are excluded in parent structures (not applicable to CDVAE and CrystalTextLLM as they are trained on data including $f$-electron elements). $^\ddagger$The stable fraction represents the percentage of DFT-verified stable structures ($E_{\text{d}} < 0.0$ eV/atom) over structures predicted to be metastable ($E_{\text{d}} < 0.1$ eV/atom) by respective surrogate models (M3GNet for CDVAE and CrystalTextLLM, CHGNet for ours, with CHGNet being more rigorous as evidenced by lower metastability rates). $^\S$We exclude structures containing $f$-electron in DFT verification while keeping the denominator as all metastable structures.
        }
    \label{tab:main_result}
\end{table*}

\subsection{Main Experimental Results}
\label{sec:CSG}

In this section, we evaluate our proposed pipeline on progressively more challenging tasks, from crystal structure generation through design to crystal structure prediction.%

\textbf{Crystal structure generation.}
We first evaluate the ability of our framework to generate stable crystal structures by optimizing decomposition energy $E_\text{d}$ as the sole objective. The LLM prompting template is detailed in \cref{appendix:prompt}. 

The generation results are reported in \cref{tab:main_result}.
Following previous work~\cite{xie2022crystal, gruver2024llmtime}, we report structural and compositional validity, which assess non-overlapping atomic radii and charge neutrality respectively. Metastability is evaluated using both CHGNet and M3GNet as surrogate models, measuring the percentage of structures with decomposition energies below 0.1~eV/atom and 0.03~eV/atom thresholds. Structures identified as metastable ($E_\text{d} < 0.1$ eV/atom) by CHGNet undergo further DFT calculations for stability assessment. 

We compare our model against two baseline models CDVAE \cite{xie2022crystal} and CrystalTextLLM \cite{gruver2024llmtime}. Among 1,479 generated structures, 76.8\% and 81.1\% are metastable based on CHGNet and M3GNet evaluations respectively, outperforming the 49.8\% metastability rate by M3GNet of the state-of-the-art CrystalTextLLM 70B model, which has a comparable model size to our base model. Under rigorous DFT validation, 31.7\% of the metastable structures remain stable, substantially improving the 10.6\% stability rate from CrystalTextLLM 70B.

However, structures containing $f$-electron elements (actinides and lanthanides, abbreviated as $f$-ele) challenge stability prediction with their strongly correlated electron interactions, which may not be adequately captured by DFT approaches under GGA and Hubbard $U$ corrections~\cite{Anisimov1997LDA+U}. These structures consistently yield lower decomposition energies ($E_\text{d}$), creating a potential computational shortcut. 
To assess this effect, we report the percentage of stable structures without $f$-ele (denoted as ``w/o $f$-ele'') among the metastable structures.
By excluding $f$-electron structures from parent selection (marked with \xmark), we improved metastability rates to 78.4\% and increased stable non-$f$-electron structures from 14.0\% to 24.6\%. This simple intervention demonstrates our framework's ability to effectively guide exploration toward diverse stable configurations, which remain unaddressed by existing methods.

While achieving better performance, our method also offers significant computational advantages. Compared to CrystalTextLLM which requires extensive fine-tuning on more than 120K structures, we achieve higher stability rates using only a few reference structures and direct LLM inference. The computational cost is primarily from structure evaluation rather than model training or fine-tuning, making our approach more accessible.
\begin{figure}
    \centering
    \begin{tabular}{@{}cc@{}}
        \begin{tabular}{@{}c@{}}
            \includegraphics[width=0.5\textwidth]{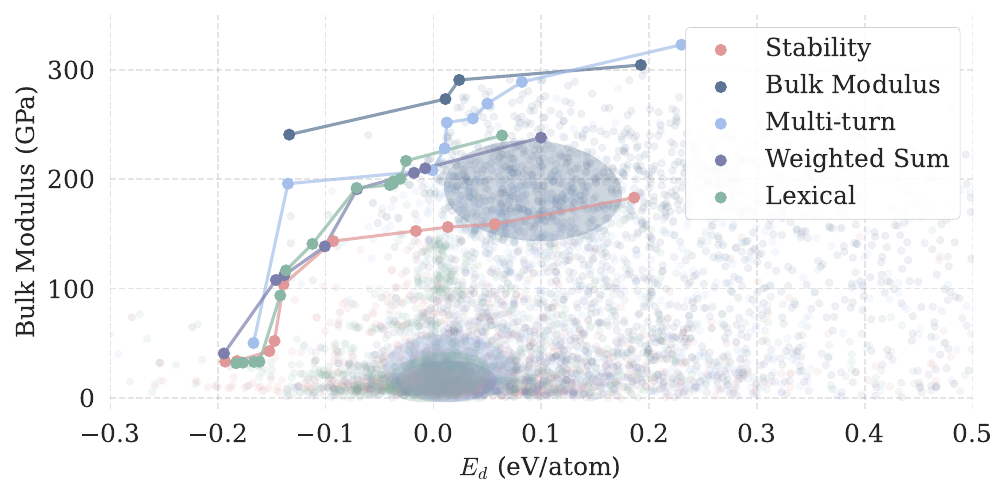} \\
             \parbox{0.5\textwidth}{\centering \small (a) Pareto frontiers}
        \end{tabular} &
        \begin{tabular}{@{}c@{}}
            \begin{tabular}{@{}c@{\hspace{-0.8cm}}c@{}}
                \includegraphics[width=0.26\textwidth]{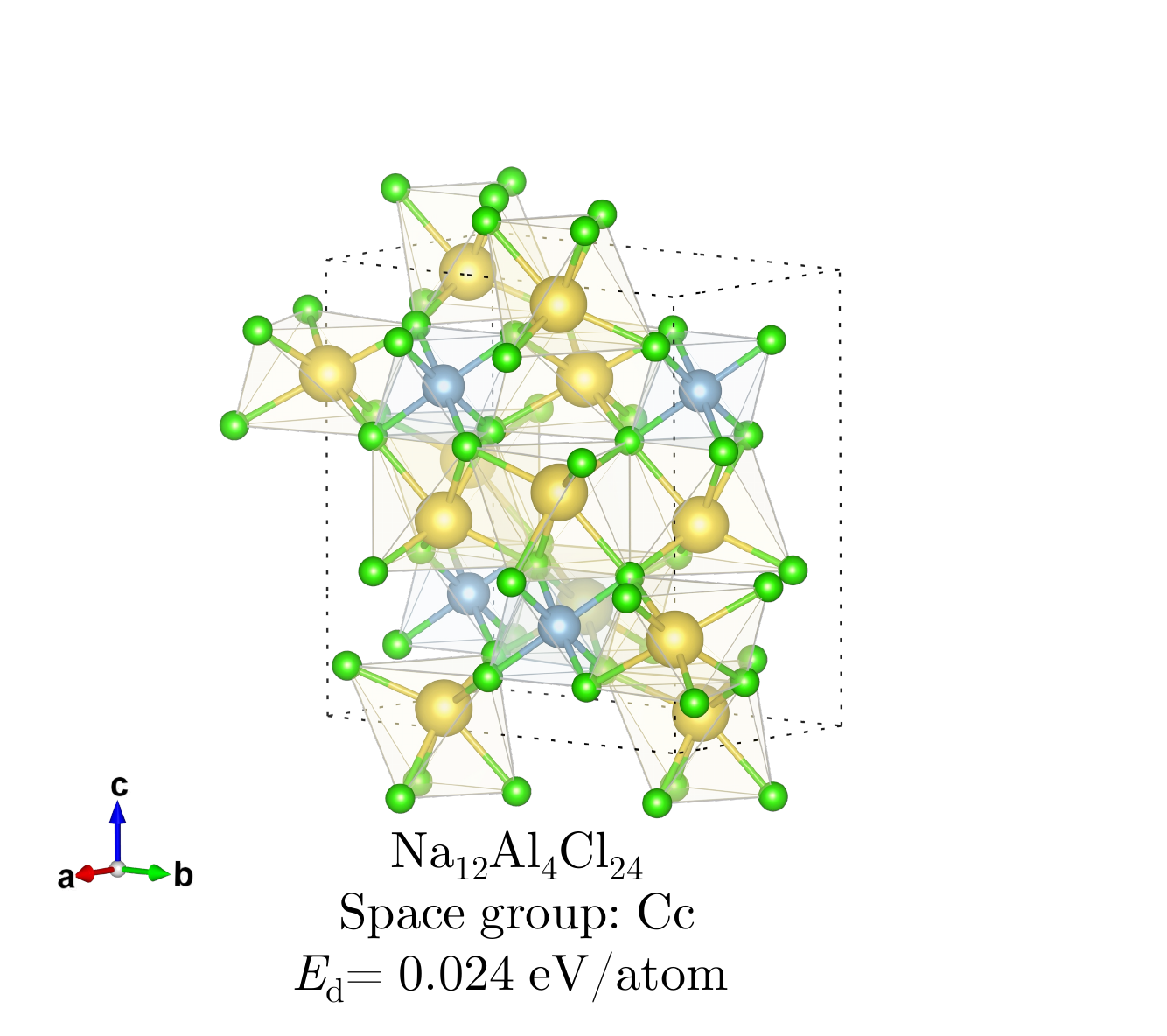} &
                \includegraphics[width=0.22\textwidth]{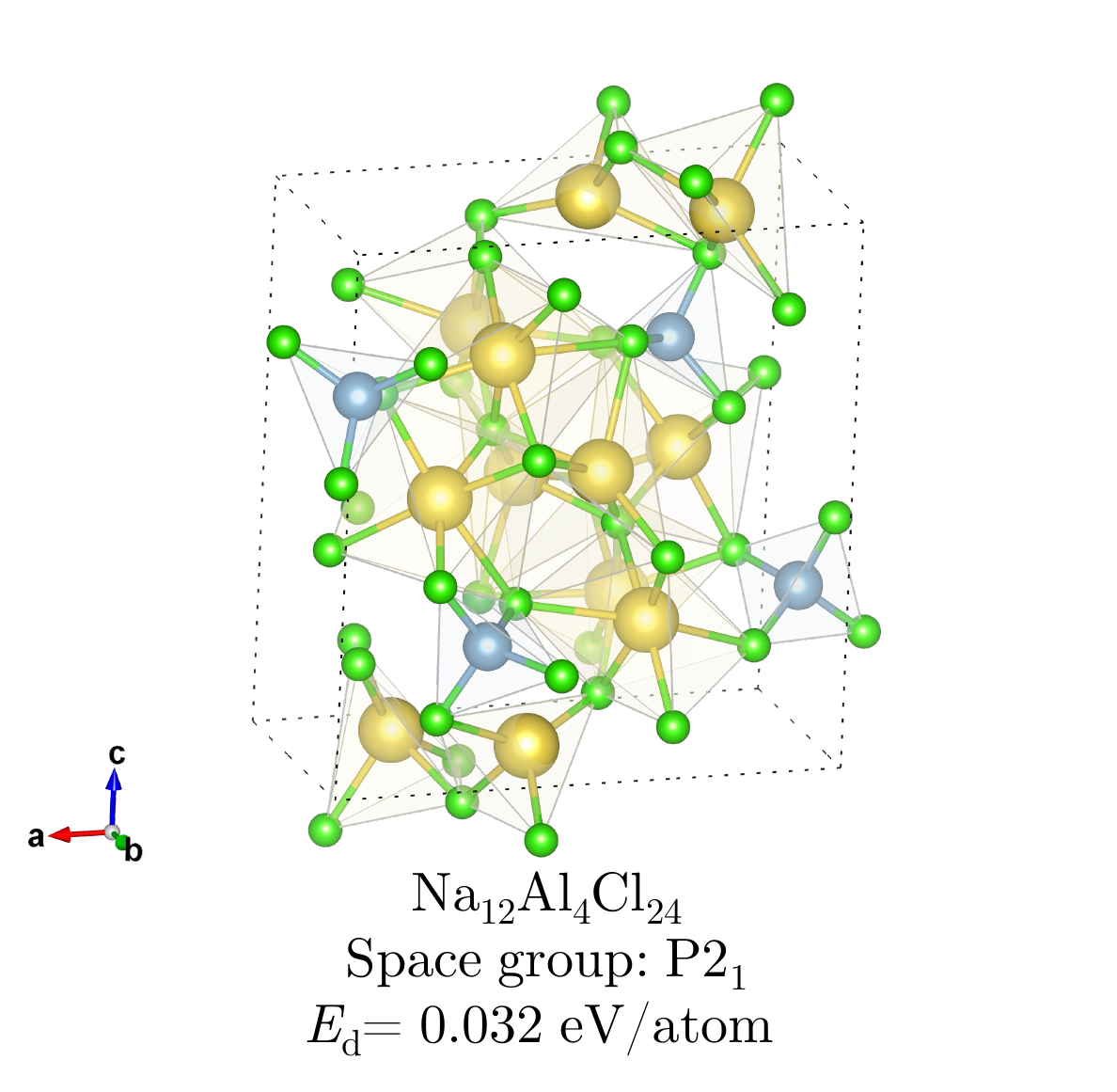} \\
            \end{tabular} \\
            \parbox{0.48\textwidth}{\centering \small (b) CSP Examples}
        \end{tabular} \\
    \end{tabular}
    \caption{(a) Pareto frontiers of bulk modulus versus decomposition energy ($E_\text{d}$) for structures optimized towards stability, bulk modulus and multi-objective (multi-turn). Ellipses indicate regions of highest structure density. (b) Examples of predicted crystal structures with composition \ce{Na3AlCl6}.}
    \label{fig:pareto_na3alcl6}
\end{figure}

\textbf{Crystal structure design.} 
We also explore multi-objective optimization by extending our framework to balance stability with desired material properties. We demonstrate this capability by alternating between optimizing stability ($E_\text{d}$) and bulk modulus in each iteration.
While this multi-objective setting naturally yields lower stability rates (57.1\% metastable with $E_\text{d} < 0.1$ eV/atom and 15.6\% DFT-verified stable structures with $f$-electron elements) compared to stability-only optimization, it enables the discovery of structures with favorable property-stability trade-offs. 

As shown in \Cref{fig:pareto_na3alcl6}(a), the Pareto frontiers under various optimization strategies converge in regions with high bulk modulus ($>200$ GPa) and metastability ($E_\text{d} \leq 0.1$ eV/atom) in the stability-property space, indicating successful discovery of potentially valuable structures that balance both objectives. The regions of highest structure density, estimated using Gaussian KDE and visualized as ellipses, reveal how optimization goals affect the distribution. Prioritizing bulk modulus shifts the density distribution toward higher mechanical strength  at the cost of increased decomposition energy.
We provide additional discussions of multi-objective optimization strategies in \cref{appendix:multi_opt}.

\textbf{Crystal structure prediction.}
We next evaluate our framework on crystal structure prediction tasks, which aim to predict stable structure (i.e. lattice and atomic coordinates) for a given composition.
As a case study, we prompt the LLM to predict polymorphs of \ce{Na3AlCl6}. For context, the Materials Project database currently contains only one structure for this composition (\texttt{mp-1111450}, Fm$\Bar{\textrm{3}}$m, $E_\text{d} = 0.142$~eV/atom), which is significantly unstable.

During the prompting process, we apply specific structural filters to select seed structures containing only three distinct elements in a 3:1:6 ratio, matching the stoichiometry of \ce{Na3AlCl6}.
From MatBench, we identified 820 structures meeting the criteria to build the initial population. Example structures proposed by the LLM for this composition are visualized in \cref{fig:pareto_na3alcl6}(b), with DFT-verified decomposition energies of 0.024 and 0.032~eV/atom respectively.
Although these predicted polymorphs remain metastable, their decomposition energies $E_\text{d}$ are significantly lower than the previously reported structure in MatBench  ($E_\text{d}$ reduced by up to 83\%), exemplifying the potential of our evolutionary pipeline for CSP applications.
We provide more successful structure prediction demonstrations including \ce{Ag6O2}, \ce{Bi2F8}, etc. in the \Cref{appendix:CSP}.

\subsection{CSG Quality Evaluation Metrics}
\label{exp:metrics}

\subsubsection{Structural and Compositional Diversity}
To better understand the effectiveness of our \ours, we analyze the diversity of generated structures and evaluate different prompting strategies, with and without fine-tuning. Additional ablation studies on factors affecting generation performance are presented in \Cref{appendix:format,appendix:hyperparameters}.

\begin{figure}
    \centering
    \begin{minipage}[b]{0.57\textwidth}
    \centering
        \includegraphics[width=0.49\linewidth]{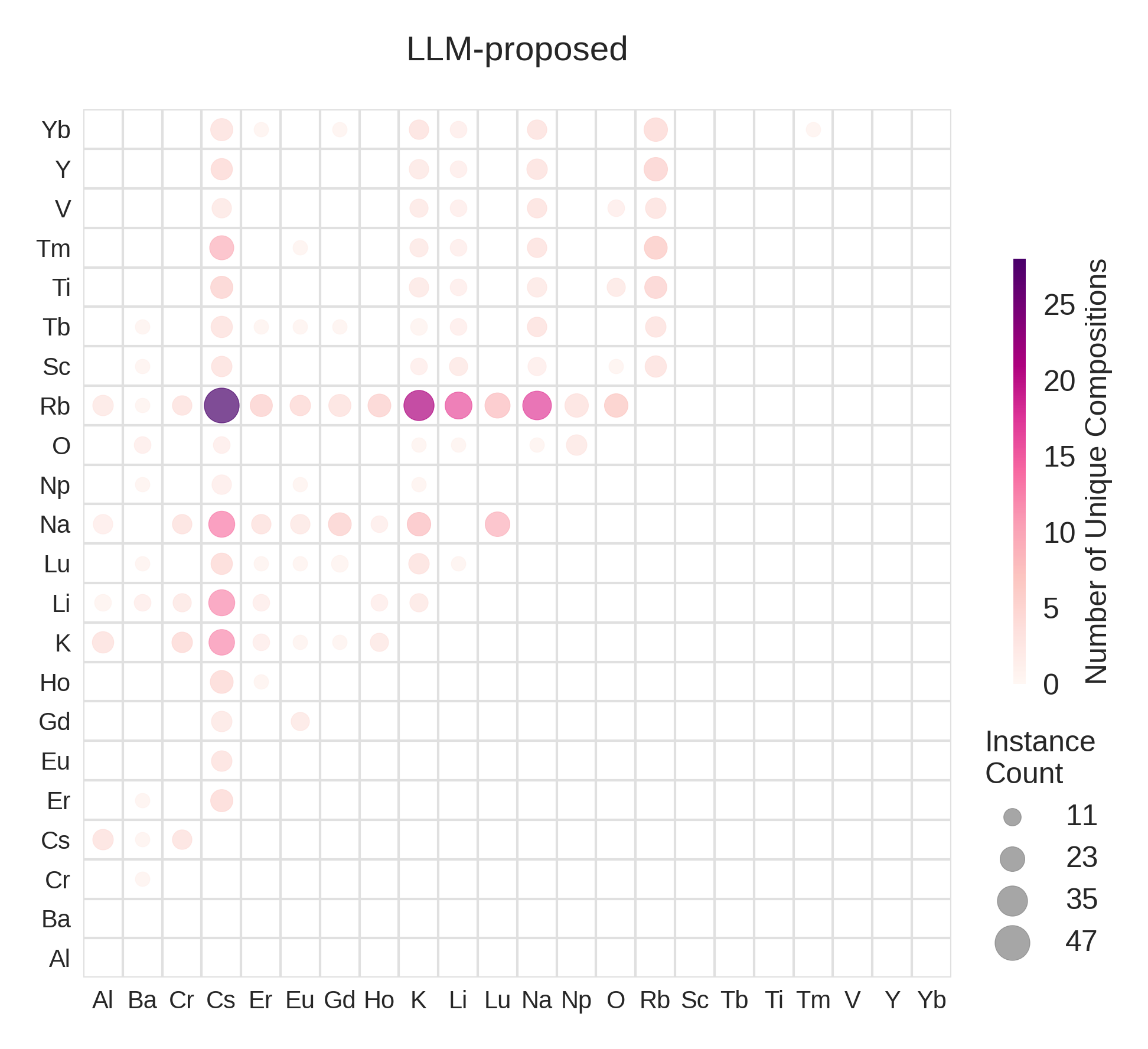}
        \includegraphics[width=0.49\linewidth]{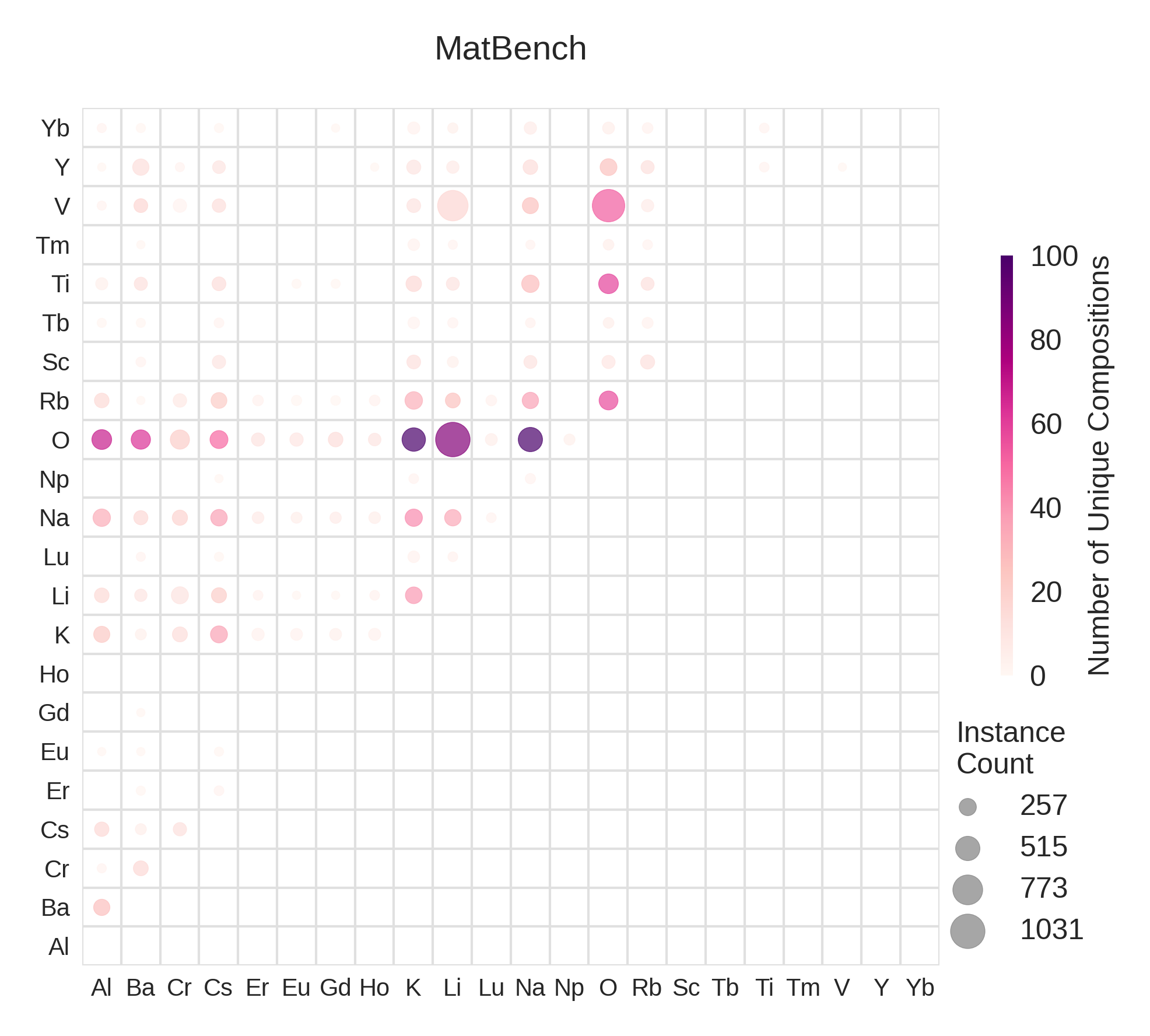}
    \caption{Element co-occurrence patterns with fluorine (F) in LLM-proposed structures (left) versus MatBench structures (right). Bubble size indicates frequency of occurrence for each element pair, while color intensity represents compositional diversity (darker indicates more unique compositions with that element pair).}
    \label{fig:combination_heatmaps}
\end{minipage}
\hfill 
\begin{minipage}[b]{0.41\textwidth}
    \centering
    \includegraphics[width=\linewidth]{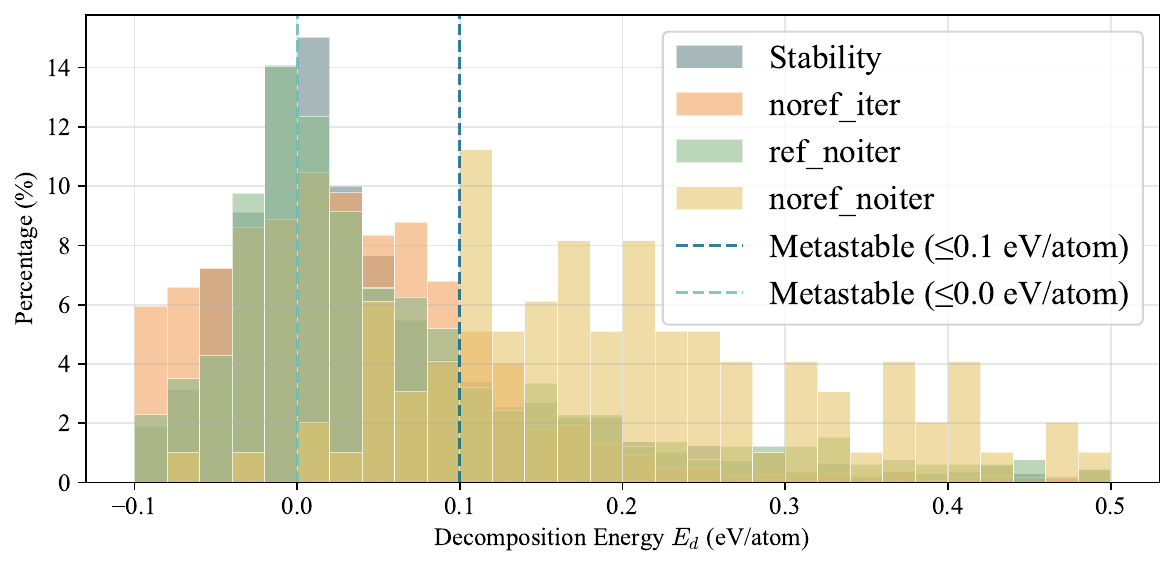}
    \caption{Decomposition energy ($E_\text{d}$) distribution comparison across experimental configurations. Vertical lines indicate metastable thresholds at 0.0 eV/atom (stable) and 0.1 eV/atom (metastable). Reference-guided approaches show more balanced distributions.}
    \label{fig:ed_distribution}
\end{minipage}
\end{figure}

To evaluate the diversity of our generated structures, we analyzed their compositional and structural characteristics by comparing LLM-proposed structures and with the $K\times P$ most stable structures from MatBench that forms the initial generation. 
Our element co-occurrence analysis reveals high compositional diversity in the LLM-proposed structures, with even the most frequent compositions appearing only twice (approximately 0.14\% of total structures). Examination of element co-occurrences with F in \cref{fig:combination_heatmaps} highlights the effectiveness of our evolutionary method in guiding structure generation toward stable F-based compounds particularly with alkali metals and transition metals.

\begin{wraptable}{r}{0.5\textwidth}
    \centering
    \small
    \SetTblrInner{rowsep=1.3pt}
    \begin{tblr}{
        colspec = {lcc},
        row{1} = {bg=gray!25},
        cell{odd[2-4]}{1-3} = {bg=gray!10}
    }
        \toprule
        \textbf{Method} & \makecell{\textbf{Stability Rate}\\ ($E_d < 0.0$ eV/atom, \%)} & \makecell{\textbf{S.U.N. Rate} \\ (\%)} \\
        \midrule
        \makecell[l]{MatLLMSearch\\(Llama 3.1-70B)} & 24.34 & 23.60 \\
        DiffCSP & 5.06 & 3.34 \\
        FlowMM & 4.65 & 2.34 \\
        \bottomrule
    \end{tblr}
    \caption{S.U.N. rate of generated structures. S.U.N. computed against MatBench reference structures used in our workflow.}
    \label{tab:sun_rate}
\end{wraptable}

The structural diversity is further evidenced by space group distribution and stability analysis for our \textbf{Stability} configuration. The generated structures demonstrates broad structural diversity with high metastability rates across multiple space groups, confirming that our evolutionary method successfully navigates toward stable regions of chemical space while maintaining diverse structural motifs across different crystallographic symmetries. 
Additional diversity and novelty evaluations and analyses are provided in \cref{appendix:diversity_novelty}.

\subsubsection{Comparison with Baseline Methods}

To ensure fair comparison with existing methods, we follow the same evaluation protocol as FlowMM~\cite{Zeni2025MatterGen}. We pre-relax all structures using CHGNet followed by density functional theory relaxation for stability verification. The results in \Cref{tab:sun_rate} demonstrate that our framework maintains higher stability rate and S.U.N. (Stable, Unique, Novel) rate than the two baselines, achieving 24.34\% stability rate and 23.60\% S.U.N. rate compared to DiffCSP and FlowMM. As S.U.N. depends on chosen reference sets and collapses multi-dimensional quality into counts, we treat it as a supplementary signal alongside multi-MLIP metastability and DFT verification. In addition, we provide comprehensive evaluation across all S.U.N. dimensions in  \Cref{appendix:diversity_novelty}: metastability rates with multiple MLIPs (CHGNet, M3GNet and Orb-v3 for CSP) and DFT-calculated stability rates for Stability; space group distributions and crystal system diversity for Uniqueness; compositional and structural novelties, elemental co-occurrence pattern shifts for Novelty. Further discussions on comparison fairness provided in \Cref{appendix:ablation} and additional generation time overhead analysis is provided in \Cref{appendix:time_overhead}.

\subsection{Ablation Analysis}
\label{sec:exp_ablation}
\begin{figure}
    \centering
    \begin{minipage}{0.47\textwidth}
        \centering
        \includegraphics[width=\linewidth]{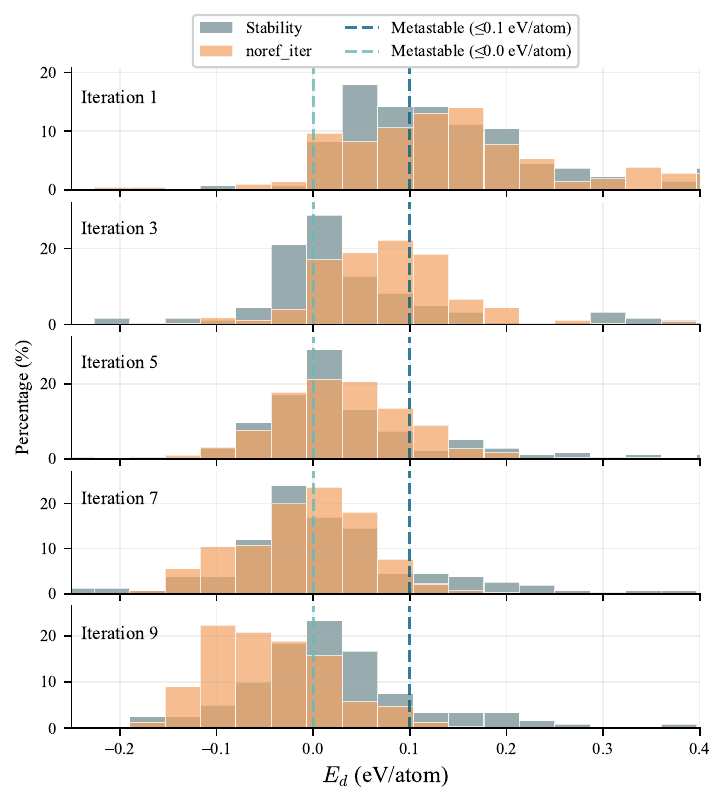}
        \small (a) $E_\text{d}$ distribution evolution
    \end{minipage}
    \hfill
    \begin{minipage}{0.52\textwidth}
        \centering
        \includegraphics[width=\linewidth]{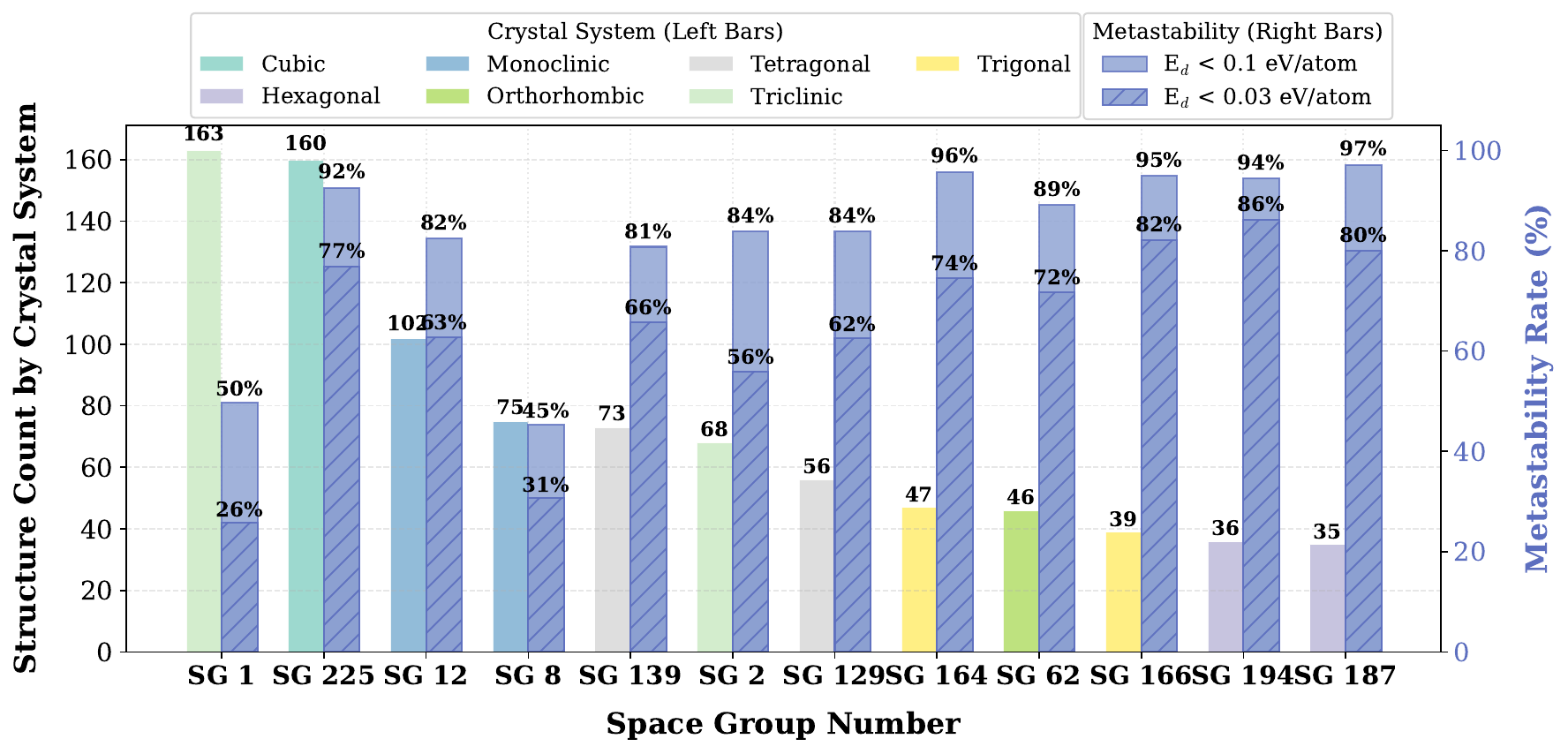}
        \small (b) \textbf{Stability}: broad space group diversity
        \includegraphics[width=\linewidth]{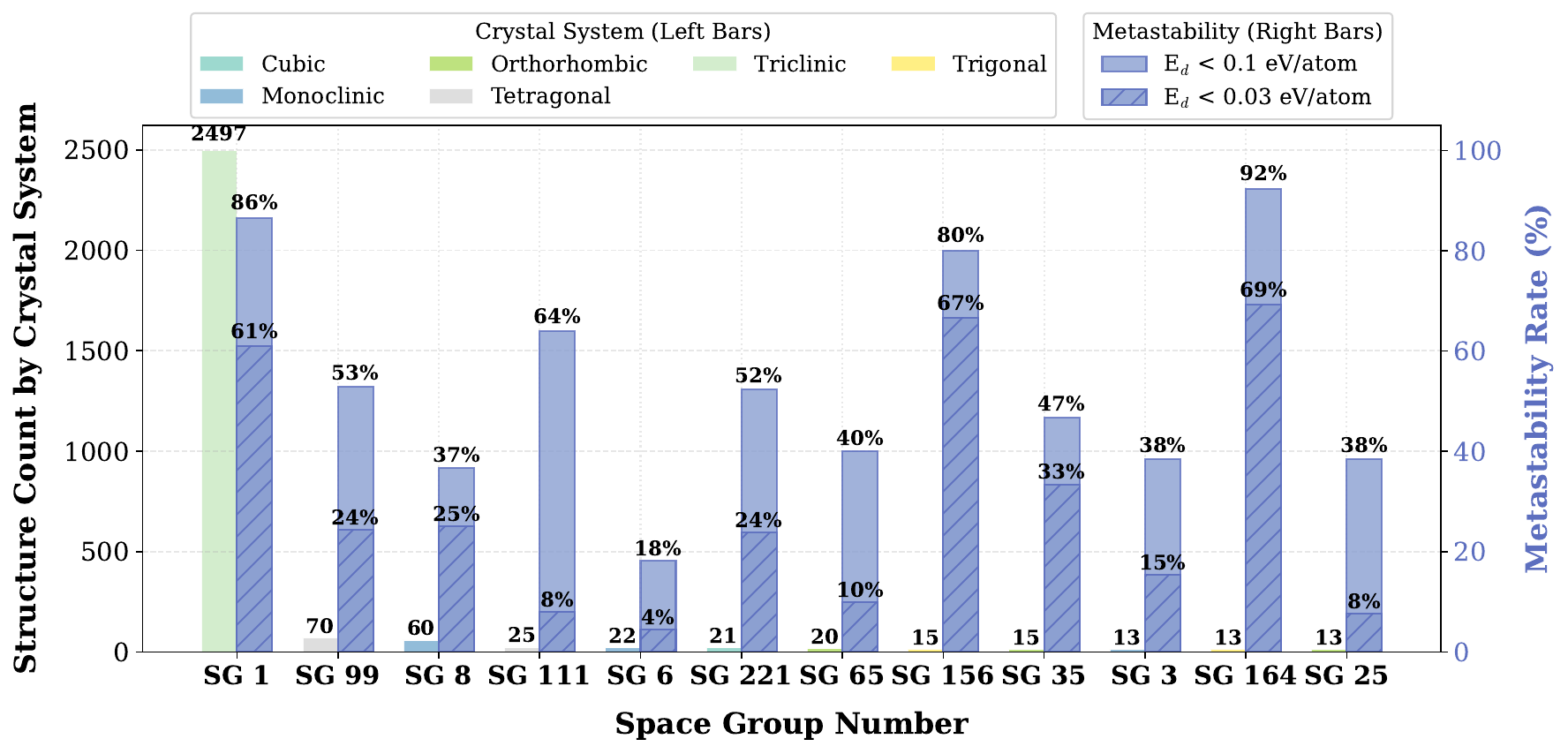}\\
        \small (c) \textbf{noref\_iter}: concentrated in space group 1
    \end{minipage}
    \caption{Ablation analysis comparing reference-guided (\textbf{Stability}) vs. reference-free (\textbf{noref\_iter}) generation. (a) $E_\text{d}$ distributions across iterations. (b,c) Space group diversity: reference structures enable broad space group diversity with high metastability (b); w/o references collapse to space group 1 (c).}
    \label{fig:ref_ablation_analysis}
\end{figure}
We evaluate four configurations to isolate the impact of reference structures and evolutionary iterations: \textbf{Stability} (full framework), \textbf{noref\_iter} (evolutionary search without references), \textbf{ref\_noiter} (single-iteration with references), and \textbf{noref\_noiter} (zero-shot). 
Decomposition energy distributions (\Cref{fig:ed_distribution}) reveal that reference-guided approaches (\textbf{Stability}, \textbf{ref\_noiter}) tightly concentrate structures near metastable thresholds, while evolutionary search alone (\textbf{noref\_iter}) shows a similar but slightly dispersed distribution. In contrast, \textbf{noref\_noiter} generates structures with substantially inferior metastability.

\textbf{Impact of evolutionary iterations.} Evolutionary iterations enable practical generation volumes: \textbf{Stability} produces 1,479 structures versus 741 for \textbf{ref\_noiter}, showing 2$\times$ higher productivity under equivalent computational budget (1,000 LLM inferences). As we investigate the distribution shift over iterations (\Cref{fig:ref_ablation_analysis}a), we observe that iterative optimization progressively shifts $E\text{d}$ distributions toward stability.

\textbf{Impact of reference structures.} While evolutionary search alone (\textbf{noref\_iter}) achieves high metastability in later iterations, it suffers structural collapse with 88\% of generated structures concentrate in triclinic space group 1 and limited exploration of other crystallographic symmetries (\Cref{fig:ref_ablation_analysis}c), indicating convergence to a single crystallographic motif. In contrast, reference-guided \textbf{Stability} configuration maintains comparable high metastability while distributing structures across 12 major space groups (\Cref{fig:ref_ablation_analysis}b). This demonstrates that reference structures prevent premature convergence while preserving thermodynamic quality.

In general, \textbf{noref\_iter} achieves high metastability but diversity collapse, \textbf{ref\_noiter} achieves structural diversity but limited volume, while \textbf{noref\_noiter} fails on both dimensions (20.7\% metastability, minimal volume). By combining reference structures with evolutionary search, \textbf{Stability} simultaneously achieves high metastability, balanced structural diversity, and practical generation scale. Additional analysis including novelty metrics, model scaling effects, and fine-tuning comparisons appears in \Cref{appendix:ablation}.

\section{Related Work}

\subsection{Language Models for Materials Science}

The increasing capabilities of LLMs have prompted materials science community to explore their potential for understanding and predicting material properties~\cite{jablonka202314}. However, benchmarking studies suggest fine-tuning LLMs over specific materials datasets is necessary to achieve performance comparable to or better than specialized graph neural networks~\cite{rubungo2024llm4mat}.
Research in crystal structure generation has developed along two main paths. \citet{flam2023language} demonstrate that autoregressive models trained from scratch with character-level tokenization can generate chemically valid crystal structures by directly tokenizing CIF files into string sequences. 
Secondly, CrystalTextLLM \cite{gruver2024llmtime} fine-tunes a pre-trained LLM (over massive texts) on generating crystalline structures with task-specific prompts. Mat2Seq \cite{yan2024invariant} converts 3D crystal structures into unique 1D sequences that preserve $\mathit{SE}(3)$ and periodic invariance for language model training. While these approaches produce valid structures, they sacrifice the general conversation capabilities of LLMs due to specialized training or fine-tuning on crystallographic data.
In parallel developments within molecular chemistry, MolLEO \cite{wang2024efficient} successfully employs pre-trained LLMs without domain-specific fine-tuning to search for small molecules. Subsequent work \cite{lu2024generative} extended this evolutionary optimization approach to more complex transition metal chemistry using advanced base LLMs with enhanced reasoning capabilities. However, these applications benefit from natural string representations for molecules (e.g., SMILES or SELFIES), which are considerably simpler than the three-dimensional representations required for crystal structures.
Our work bridges this gap by adapting the evolutionary approach to the more complex domain of crystal structures without requiring fine-tuning.

\subsection{Generative Models for Materials Discovery}

Besides autoregressive language models, various generative models including variational autoencoders, diffusion models, and flow models have emerged as promising solutions for crystal structure generation. 
Early work proposes generative crystal structures using variational autoencoders that represent crystal structures as 3D voxels \cite{noh2019inverse,court20203}. CDVAE first proposes to generate crystal structures with a score-based generative (diffusion) model and optimize crystal structure properties through gradient-based optimization in the latent space \cite{xie2022crystal}.
This approach has been extended in several directions: \citet{jiao2024crystal} developed Riemannian diffusion models to better handle periodic coordinates, \citet{Zeni2025MatterGen} scaled the approach to encompass elements across the entire periodic table with various design criteria, and \citet{dai2024inpainting} applied it to crystal inpainting tasks. Most recently, \citet{sriram2024flowllm} introduced Riemannian flow matching models to better address periodic boundary conditions with improved performance. \citet{yang2024generative} explore the synergy between language and generative models by leveraging LLMs to propose chemical formulae under design constraints before feeding them to a diffusion model.

\subsection{Classical Substitution Methods}
Classical substitution techniques generate candidates by replacing elements within known prototypes according to heuristic rules or ionic radii/valence compatibility. Such methods are efficient and ensure validity within predefined chemical spaces, but they are limited by the expressivity of the rule set and rarely propose nuanced edits such as partial substitutions or motif-level reorganizations. Our training-free LLM-based framework differs in goal and mechanism: by conditioning on multiple parents and free-form text structure strings, it proposes chemically coherent edits beyond strict substitution, while evolutionary selection guides validity and stability. We regard substitution as a complementary initialization strategy rather than a direct baseline to our broader search capability.

\section{Conclusion}

In this paper, we present an evolutionary workflow for computational materials discovery, encompassing crystal structure generation, prediction, and objective-based optimization. We demonstrate that a pre-trained LLM trained on general text can identify a higher proportion of (meta)stable materials compared to state-of-the-art generative models specifically trained on materials datasets. These findings suggest that LLMs inherently function as effective crystal structure generators, with both compositional and structural information naturally embedded within their text inference capabilities. 
In conclusion, our method complements existing structure discovery techniques by providing refined optimization capabilities while maintaining versatility in addressing various optimization objectives, offering an efficient approach for high-throughput materials discovery.

Looking forward, a natural extension of this work would be synthesis prediction based on the evolutionary method. Improved machine learning interatomic potentials will complement this process, as discussed in \cref{appendix:MLIP}. Such development would benefit from integration with high-quality experimental data from automated, high-throughput experiments, bridging the gap between computational predictions and experimental synthesis, which would accelerate high-throughput materials discovery.

\bibliographystyle{plainnat} 
\bibliography{bibliography}

\begin{thebibliography}{62}
\providecommand{\natexlab}[1]{#1}
\providecommand{\url}[1]{\texttt{#1}}
\expandafter\ifx\csname urlstyle\endcsname\relax
  \providecommand{\doi}[1]{doi: #1}\else
  \providecommand{\doi}{doi: \begingroup \urlstyle{rm}\Url}\fi

\bibitem[Ani()]{Anisimov1997LDA+U}
Hubbard-corrected dft energy functionals: The lda + u description of correlated systems.

\bibitem[Achiam et~al.(2023)Achiam, Adler, Agarwal, et~al.]{achiam2023gpt}
Josh Achiam, Steven Adler, Sandhini Agarwal, et~al.
\newblock {GPT-4 Technical Report}.
\newblock \emph{arXiv.org}, 2023.

\bibitem[Allahyari and Oganov(2020)]{allahyari2020coevolutionary}
Zahed Allahyari and Artem~R. Oganov.
\newblock {Coevolutionary Search for Optimal Materials in the Space of All Possible Compounds}.
\newblock \emph{npj Comput. Mater.}, 2020.

\bibitem[Antunes et~al.(2023)Antunes, Butler, and Grau‐Crespo]{Antunes2023CrystalSG}
Luis~M. Antunes, Keith~T. Butler, and Ricardo Grau‐Crespo.
\newblock {Crystal Structure Generation with Autoregressive Large Language Modeling}.
\newblock \emph{Nat. Commun.}, 2023.

\bibitem[Bagayoko(2014)]{bagayoko2014dft}
Diola Bagayoko.
\newblock {Understanding Density Functional Theory (DFT) and Completing It in Practice}.
\newblock \emph{AIP Adv.}, 2014.

\bibitem[Barroso-Luque et~al.(2024)Barroso-Luque, Shuaibi, Fu, et~al.]{barroso2024open}
Luis Barroso-Luque, Muhammed Shuaibi, Xiang Fu, et~al.
\newblock {Open Materials 2024 (OMAT24) Inorganic Materials Dataset and Models}.
\newblock \emph{arXiv.org}, 2024.

\bibitem[Batatia et~al.(2023)Batatia, Benner, Chiang, et~al.]{batatia2023foundation}
Ilyes Batatia, Philipp Benner, Yuan Chiang, et~al.
\newblock {A Foundation Model for Atomistic Materials Chemistry}.
\newblock \emph{arXiv.org}, 2023.

\bibitem[Batzner et~al.(2022)Batzner, Musaelian, Sun, et~al.]{batzner2022nequip}
Simon Batzner, Albert Musaelian, Lixin Sun, et~al.
\newblock {E(3)-Equivariant Graph Neural Networks for Data-Efficient and Accurate Interatomic Potentials}.
\newblock \emph{Nat. Commun.}, 2022.

\bibitem[Bitzek et~al.(2006)Bitzek, Koskinen, G{\"a}hler, et~al.]{Bitzek2006_FIRE}
Erik Bitzek, Pekka Koskinen, Franz G{\"a}hler, et~al.
\newblock {Structural Relaxation Made Simple}.
\newblock \emph{Phys. Rev. Lett.}, 2006.

\bibitem[Boiko et~al.(2023)Boiko, MacKnight, Kline, and Gomes]{Boiko:2023ky}
Daniil~A. Boiko, Robert MacKnight, Ben Kline, and Gabe Gomes.
\newblock {Autonomous Chemical Research with Large Language Models}.
\newblock \emph{Nature}, 624\penalty0 (7992):\penalty0 570--578, December 2023.

\bibitem[Bran et~al.(2024)Bran, Cox, Schilter, Baldassari, White, and Schwaller]{Bran:2024dw}
Andres~M. Bran, Sam Cox, Oliver Schilter, Carlo Baldassari, Andrew~D. White, and Philippe Schwaller.
\newblock {Augmenting Large Language Models with Chemistry Tools}.
\newblock \emph{Nat. Mach. Intell.}, 2024.

\bibitem[Chen and Ong(2022)]{chen2022m3gnet}
Chi Chen and Shyue Ong.
\newblock {A Universal Graph Deep Learning Interatomic Potential for the Periodic Table}.
\newblock \emph{Nat. Comput. Sci.}, 2022.

\bibitem[Cheng(2024)]{cheng2024cartesian}
Bingqing Cheng.
\newblock {Cartesian Atomic Cluster Expansion for Machine Learning Interatomic Potentials}.
\newblock \emph{npj Comput. Mater.}, 2024.

\bibitem[Court et~al.(2020)Court, Yildirim, Jain, et~al.]{court20203}
Callum~J. Court, Batuhan Yildirim, Apoorv Jain, et~al.
\newblock {3-D Inorganic Crystal Structure Generation and Property Prediction via Representation Learning}.
\newblock \emph{J. Chem. Inf. Model.}, 2020.

\bibitem[Curtis et~al.(2018)Curtis, Li, Rose, et~al.]{curtis2018GAtor}
Farren Curtis, Xiayue Li, Timothy Rose, et~al.
\newblock {GAtor: A First-Principles Genetic Algorithm for Molecular Crystal Structure Prediction}.
\newblock \emph{J. Chem. Theory Comput.}, 2018.

\bibitem[Dai et~al.(2024)Dai, Zhong, Deng, et~al.]{dai2024inpainting}
Xinzhe Dai, Peichen Zhong, Bowen Deng, et~al.
\newblock {Inpainting Crystal Structure Generations with Score-Based Denoising}.
\newblock In \emph{ICML Workshop on AI for Science}, 2024.

\bibitem[Deng et~al.(2023)Deng, Zhong, Jun, et~al.]{deng2023chgnet}
Bowen Deng, Peichen Zhong, KyuJung Jun, et~al.
\newblock {CHGNet as a Pretrained Universal Neural Network Potential for Charge-Informed Atomistic Modelling}.
\newblock \emph{Nat. Mach. Intell.}, 2023.

\bibitem[Du et~al.(2023{\natexlab{a}})Du, Wang, Feng, et~al.]{du2024new}
Yuanqi Du, Limei Wang, Dieqiao Feng, et~al.
\newblock {A New Perspective on Building Efficient and Expressive 3D Equivariant Graph Neural Networks}.
\newblock \emph{NeurIPS}, 2023{\natexlab{a}}.

\bibitem[Du et~al.(2023{\natexlab{b}})Du, Wang, Huang, et~al.]{du2023m}
Yuanqi Du, Yingheng Wang, Yining Huang, et~al.
\newblock {M$^2$Hub: Unlocking the Potential of Machine Learning for Materials Discovery}.
\newblock \emph{NeurIPS}, 2023{\natexlab{b}}.

\bibitem[Dunn et~al.(2020)Dunn, Wang, Ganose, et~al.]{dunn2020benchmarking}
Alexander Dunn, Qi~Wang, Alex Ganose, et~al.
\newblock {Benchmarking Materials Property Prediction Methods: The Matbench Test Set and Automatminer Reference Algorithm}.
\newblock \emph{npj Comput. Mater.}, 2020.

\bibitem[Eremin et~al.(2023)Eremin, Humonen, Kazakov, et~al.]{eremin2023gnn}
Roman Eremin, Innokentiy Humonen, Alexey Kazakov, et~al.
\newblock {Graph Neural Networks for Predicting Structural Stability of Cd- and Zn-doped $\lambda$-CsPbI$_3$}.
\newblock \emph{Comput. Mater. Sci.}, 2023.

\bibitem[Flam-Shepherd and Aspuru-Guzik(2023)]{flam2023language}
Daniel Flam-Shepherd and Al{'a}n Aspuru-Guzik.
\newblock {Language Models Can Generate Molecules, Materials, and Protein Binding Sites Directly in Three Dimensions as XYZ, CIF, and PDB Files}.
\newblock \emph{arXiv.org}, 2023.

\bibitem[Fu et~al.(2023)Fu, Wei, Song, et~al.]{Fu_2023}
Nihang Fu, Lai Wei, Yuqi Song, et~al.
\newblock {Material Transformers: Deep Learning Language Models for Generative Materials Design}.
\newblock \emph{Mach. Learn.: Sci. Technol.}, 2023.

\bibitem[Grattafiori et~al.(2024)Grattafiori, Dubey, Jauhri, et~al.]{grattafiori2024llama3herdmodels}
Aaron Grattafiori, Abhimanyu Dubey, Abhinav Jauhri, et~al.
\newblock {The Llama 3 Herd of Models}.
\newblock \emph{arXiv.org}, 2024.

\bibitem[Gruver et~al.(2024)Gruver, Sriram, Madotto, et~al.]{gruver2024llmtime}
Nate Gruver, Anuroop Sriram, Andrea Madotto, et~al.
\newblock {Fine-Tuned Language Models Generate Stable Inorganic Materials as Text}.
\newblock In \emph{ICLR}, 2024.

\bibitem[Guo et~al.(2023)Guo, Guo, Nan, Liang, Guo, Chawla, Wiest, and Zhang]{Guo:2023wn}
Taicheng Guo, Kehan Guo, Bozhao Nan, Zhenwen Liang, Zhichun Guo, Nitesh~V. Chawla, Olaf Wiest, and Xiangliang Zhang.
\newblock {What Can Large Language Models Do in Chemistry? A Comprehensive Benchmark on Eight Tasks}.
\newblock In \emph{NeurIPS}, 2023.

\bibitem[Heiles and Johnston(2013)]{heiles2013global}
Sven Heiles and Roy~L. Johnston.
\newblock {Global Optimization of Clusters Using Electronic Structure Methods}.
\newblock \emph{Int. J. Quantum Chem.}, 2013.

\bibitem[Jablonka et~al.(2023)Jablonka, Ai, Al-Feghali, et~al.]{jablonka202314}
Kevin~M. Jablonka, Qianxiang Ai, Alexander Al-Feghali, et~al.
\newblock {14 Examples of How LLMs Can Transform Materials Science and Chemistry: A Reflection on a Large Language Model Hackathon}.
\newblock \emph{Digit. Discov.}, 2023.

\bibitem[Jain et~al.(2011)Jain, Hautier, Ong, et~al.]{Jain2011_GGAU}
Anubhav Jain, Geoffroy Hautier, Shyue~P. Ong, et~al.
\newblock {Formation Enthalpies by Mixing GGA and GGA + U Calculations}.
\newblock \emph{Phys. Rev. B}, 2011.

\bibitem[Jain et~al.(2013)Jain, Ong, Hautier, et~al.]{Jain2013MP}
Anubhav Jain, Shyue~P. Ong, Geoffroy Hautier, et~al.
\newblock {Commentary: The Materials Project: A Materials Genome Approach to Accelerating Materials Innovation}.
\newblock \emph{APL Mater.}, 2013.

\bibitem[Jiao et~al.(2024)Jiao, Huang, Lin, et~al.]{jiao2024crystal}
Rui Jiao, Wenbing Huang, Peijia Lin, et~al.
\newblock {Crystal Structure Prediction by Joint Equivariant Diffusion}.
\newblock \emph{NeurIPS}, 2024.

\bibitem[Johnston(2003)]{johnston2003evolving}
Roy~L. Johnston.
\newblock {Evolving Better Nanoparticles: Genetic Algorithms for Optimising Cluster Geometries}.
\newblock \emph{Dalton Trans.}, 2003.

\bibitem[Kadan et~al.(2023)Kadan, Ryczko, Wildman, et~al.]{kadan2023GAmuza}
Amit Kadan, Kevin Ryczko, Andrew Wildman, et~al.
\newblock {Accelerated Organic Crystal Structure Prediction with Genetic Algorithms and Machine Learning}.
\newblock \emph{J. Chem. Theory Comput.}, 2023.

\bibitem[Kresse and Furthm{\"u}ller(1996)]{Kresse1996_VASP}
G.~Kresse and J.~Furthm{\"u}ller.
\newblock {Efficient Iterative Schemes for Ab Initio Total-Energy Calculations Using a Plane-Wave Basis Set}.
\newblock \emph{Phys. Rev. B}, 1996.

\bibitem[Kresse and Joubert(1999)]{Kresse1999_PAW}
G.~Kresse and D.~Joubert.
\newblock {From Ultrasoft Pseudopotentials to the Projector Augmented-Wave Method}.
\newblock \emph{Phys. Rev. B}, 1999.

\bibitem[Liao et~al.(2024)Liao, Wood, Das, et~al.]{equiformer_v2}
Yi-Lun Liao, Brandon Wood, Abhishek Das, et~al.
\newblock {EquiformerV2: Improved Equivariant Transformer for Scaling to Higher-Degree Representations}.
\newblock In \emph{ICLR}, 2024.

\bibitem[L{'o}pez-Zorrilla et~al.(2023)L{'o}pez-Zorrilla, Aretxabaleta, Yeu, et~al.]{lopez2023aenet}
Jon L{'o}pez-Zorrilla, Xabier~M. Aretxabaleta, In~Won Yeu, et~al.
\newblock {{\ae}net-PyTorch: A GPU-Supported Implementation for Machine Learning Atomic Potentials Training}.
\newblock \emph{J. Chem. Phys.}, 2023.

\bibitem[Lu et~al.(2024)Lu, Song, Zhao, et~al.]{lu2024generative}
Jieyu Lu, Zhangde Song, Qiyuan Zhao, et~al.
\newblock {Generative Design of Functional Metal Complexes Utilizing the Internal Knowledge of Large Language Models}.
\newblock \emph{arXiv.org}, 2024.

\bibitem[Merchant et~al.(2023)Merchant, Batzner, Schoenholz, et~al.]{merchant2023GNoME}
Amil Merchant, Simon Batzner, Samuel Schoenholz, et~al.
\newblock {Scaling Deep Learning for Materials Discovery}.
\newblock \emph{Nature}, 2023.

\bibitem[Noh et~al.(2019)Noh, Kim, Stein, et~al.]{noh2019inverse}
Juhwan Noh, Jaehoon Kim, Helge~S. Stein, et~al.
\newblock {Inverse Design of Solid-State Materials via a Continuous Representation}.
\newblock \emph{Matter}, 2019.

\bibitem[Oganov and Glass(2006)]{oganov2006evolutionary}
Artem Oganov and Colin Glass.
\newblock Crystal structure prediction using ab initio evolutionary techniques: Principles and applications.
\newblock \emph{The Journal of chemical physics}, 124:\penalty0 244704, 07 2006.
\newblock \doi{10.1063/1.2210932}.

\bibitem[Park et~al.(2024)Park, Kim, Hwang, et~al.]{park_scalable_2024}
Yutack Park, Jaesun Kim, Seungwoo Hwang, et~al.
\newblock {Scalable Parallel Algorithm for Graph Neural Network Interatomic Potentials in Molecular Dynamics Simulations}.
\newblock \emph{J. Chem. Theory Comput.}, 2024.

\bibitem[Peeperkorn et~al.(2024)Peeperkorn, Kouwenhoven, Brown, et~al.]{Peeperkorn2024IsTT}
Max Peeperkorn, Tom Kouwenhoven, Daniel~G. Brown, et~al.
\newblock {Is Temperature the Creativity Parameter of Large Language Models?}
\newblock In \emph{ICCC}, 2024.

\bibitem[Perdew et~al.(1996)Perdew, Burke, and Ernzerhof]{perdew1996_GGA}
John~P. Perdew, Kieron Burke, and Matthias Ernzerhof.
\newblock {Generalized Gradient Approximation Made Simple}.
\newblock \emph{Phys. Rev. Lett.}, 1996.

\bibitem[Rafailov et~al.(2024)Rafailov, Sharma, Mitchell, et~al.]{rafailov2024direct}
Rafael Rafailov, Archit Sharma, Eric Mitchell, et~al.
\newblock {Direct Preference Optimization: Your Language Model Is Secretly a Reward Model}.
\newblock \emph{NeurIPS}, 2024.

\bibitem[Rhodes et~al.(2025)Rhodes, Vandenhaute, Šimkus, Gin, Godwin, Duignan, and Neumann]{rhodes2025orbv3}
Benjamin Rhodes, Sander Vandenhaute, Vaidotas Šimkus, James Gin, Jonathan Godwin, Tim Duignan, and Mark Neumann.
\newblock Orb-v3: atomistic simulation at scale, 2025.
\newblock URL \url{https://arxiv.org/abs/2504.06231}.

\bibitem[Rubungo et~al.(2024)Rubungo, Li, Hattrick-Simpers, et~al.]{rubungo2024llm4mat}
Andre~N. Rubungo, Kangming Li, Jason Hattrick-Simpers, et~al.
\newblock {LLM4Mat-Bench: Benchmarking Large Language Models for Materials Property Prediction}.
\newblock \emph{arXiv.org}, 2024.

\bibitem[Shoghi et~al.(2024)Shoghi, Kolluru, Kitchin, et~al.]{shoghi2023molecules}
Nima Shoghi, Adeesh Kolluru, John~R. Kitchin, et~al.
\newblock {From Molecules to Materials: Pre-Training Large Generalizable Models for Atomic Property Prediction}.
\newblock In \emph{ICLR}, 2024.

\bibitem[Sriram et~al.(2024)Sriram, Miller, Chen, et~al.]{sriram2024flowllm}
Anuroop Sriram, Benjamin~K. Miller, Ricky T.~Q. Chen, et~al.
\newblock {FlowLLM: Flow Matching for Material Generation with Large Language Models as Base Distributions}.
\newblock In \emph{NeurIPS}, 2024.

\bibitem[Sun et~al.(2016)Sun, Dacek, Ong, et~al.]{sun2016dft}
Wenhao Sun, Stephen~T. Dacek, Shyue~P. Ong, et~al.
\newblock {The Thermodynamic Scale of Inorganic Crystalline Metastability}.
\newblock \emph{Sci. Adv.}, 2016.

\bibitem[Wang et~al.(2021)Wang, Kingsbury, McDermott, et~al.]{wang2021dft}
Amanda Wang, Ryan Kingsbury, Matthew McDermott, et~al.
\newblock {A Framework for Quantifying Uncertainty in DFT Energy Corrections}.
\newblock \emph{Sci. Rep.}, 2021.

\bibitem[Wang et~al.(2024)Wang, Skreta, Ser, et~al.]{wang2024efficientevolutionarysearchchemical}
Haorui Wang, Marta Skreta, Cher-Tian Ser, et~al.
\newblock Efficient evolutionary search over chemical space with large language models.
\newblock 2024.
\newblock URL \url{https://arxiv.org/abs/2406.16976}.

\bibitem[Wang et~al.(2025)Wang, Skreta, Ser, et~al.]{wang2024efficient}
Haorui Wang, Marta Skreta, Cher-Tian Ser, et~al.
\newblock {Efficient Evolutionary Search over Chemical Space with Large Language Models}.
\newblock In \emph{ICLR}, 2025.

\bibitem[Wen et~al.(2024)Wen, Horton, Munro, et~al.]{wen2024equivariant}
Mingjian Wen, Matthew~K. Horton, Jason~M. Munro, et~al.
\newblock {An Equivariant Graph Neural Network for the Elasticity Tensors of All Seven Crystal Systems}.
\newblock \emph{Digit. Discov.}, 2024.

\bibitem[Xie et~al.(2022)Xie, Fu, Ganea, et~al.]{xie2022crystal}
Tian Xie, Xiang Fu, Octavian-Eugen Ganea, et~al.
\newblock {Crystal Diffusion Variational Autoencoder for Periodic Material Generation}.
\newblock In \emph{ICLR}, 2022.

\bibitem[Yan et~al.(2024)Yan, Li, Ling, Ashen, Edwards, Arroyave, Zitnik, Ji, Qian, Qian, and Ji]{yan2024invariant}
Keqiang Yan, Xiner Li, Hongyi Ling, Kenna Ashen, Carl Edwards, Raymundo Arroyave, Marinka Zitnik, Heng Ji, Xiaofeng Qian, Xiaoning Qian, and Shuiwang Ji.
\newblock Invariant tokenization of crystalline materials for language model enabled generation.
\newblock In \emph{The Thirty-eighth Annual Conference on Neural Information Processing Systems}, 2024.
\newblock URL \url{https://openreview.net/forum?id=18FGRNd0wZ}.

\bibitem[Yang et~al.(2024{\natexlab{a}})Yang, Hu, Zhou, et~al.]{yang2024mattersim}
Han Yang, Chenxi Hu, Yichi Zhou, et~al.
\newblock {MatterSim: A Deep Learning Atomistic Model Across Elements, Temperatures and Pressures}.
\newblock \emph{arXiv.org}, 2024{\natexlab{a}}.

\bibitem[Yang et~al.(2024{\natexlab{b}})Yang, Batzner, Gao, Aykol, Gaunt, McMorrow, Rezende, Schuurmans, Mordatch, and Cubuk]{yang2024generative}
Sherry Yang, Simon Batzner, Ruiqi Gao, Muratahan Aykol, Alexander~L Gaunt, Brendan McMorrow, Danilo~Jimenez Rezende, Dale Schuurmans, Igor Mordatch, and Ekin~Dogus Cubuk.
\newblock Generative hierarchical materials search.
\newblock In \emph{The Thirty-eighth Annual Conference on Neural Information Processing Systems}, 2024{\natexlab{b}}.
\newblock URL \url{https://openreview.net/forum?id=PsPR4NOiRC}.

\bibitem[Yin et~al.(2025)Yin, Wang, Du, et~al.]{yin2025alphanetscalinglocalframebased}
Bangchen Yin, Jiaao Wang, Weitao Du, et~al.
\newblock {AlphaNet: Scaling Up Local Frame-Based Atomistic Foundation Model}.
\newblock \emph{arXiv.org}, 2025.

\bibitem[Zeni et~al.(2025)Zeni, Pinsler, Z{"u}gner, et~al.]{Zeni2025MatterGen}
Claudio Zeni, Robert Pinsler, Daniel Z{"u}gner, et~al.
\newblock {A Generative Model for Inorganic Materials Design}.
\newblock \emph{Nature}, 2025.

\bibitem[Zhang et~al.(2024)Zhang, Bi, Dai, et~al.]{zhang2024dpa1}
Duo Zhang, Hangrui Bi, Fu-Zhi Dai, et~al.
\newblock {Pretraining of Attention-Based Deep Learning Potential Model for Molecular Simulation}.
\newblock \emph{npj Comput. Math.}, 2024.

\bibitem[Zhang et~al.(2021)Zhang, Wang, Car, et~al.]{zhang2021phase}
Linfeng Zhang, Han Wang, Roberto Car, et~al.
\newblock {Phase Diagram of a Deep Potential Water Model}.
\newblock \emph{Phys. Rev. Lett.}, 2021.

\end{thebibliography}

\newpage
\appendix
\clearpage
\appendix
\onecolumn
\appendixprefix
\vspace*{1pt}
\begin{center}
	{\Large \textbf{Supplementary Material for \ours}}
\end{center}

\section{Experimental Details of Population Initialization}
\label{appendix:exp_details}

The retrieval set $\mathcal{R}$ of stable structures are sampled from known stable structures--Matbench-bandgap dataset ~\cite{dunn2020benchmarking}, which consists of 106,113 crystal structures in total. To initialize the parent structures for the first iteration, we calculated the decomposition energy for each structure with CHGNet. For CSG task, we removed binary compounds and structures with high-order compositions, i.e., retaining candidate structures with 3 to 6 elements. For CSP tasks, we filtered the seed structures to find those matching a desired compositional pattern. In addition, we applied de-duplication by composition to the candidate structures.
The analysis of how sampling rule of the extra pool affect the performance is provided in \Cref{sec:exp_ablation}. To further enhance the structure generation, we envision future work that could explore how structures can be ensembled to form a larger candidate pool for parent selection.

\section{Reproducibility}

The crystal structures generated by \ours can be downloaded \href{https://drive.google.com/file/d/1y88pdUFKmmFxFWW4MnwCVrKSjekmOXBh/view?usp=sharing}{here}. 

The implementation of our evolutionary search pipeline is available \href{https://github.com/JingruG/MatLLMSearch}{here}.

\section{Prompt for CSG}
\label{appendix:prompt}
\begin{center}
\fbox{\parbox{0.85\textwidth}{
\itshape
You are an expert material scientist. Your task is to propose hypotheses for \{reproduction\_size\} new materials with valid stable structures and compositions. No isolated or overlapped atoms are allowed.\\[0.5em]

The proposed new materials can be a modification or combination of the base materials given below.\\[0.5em]

Format requirements:\\[0.5em]
\hspace*{2em}1. Each proposed structure must be formatted in JSON with the following structure:\\[0.5em]
{\tt\hspace*{4em}\{\{\\
\hspace*{6em}"i": \{\{\\
\hspace*{8em}"formula": "composition\_formula",\\
\hspace*{8em}"POSCAR": "POSCAR\_format\_string"\\
\hspace*{6em}\}\}\\
\hspace*{4em}\}\}}\\[0.5em]
\hspace*{2em}2. Use proper JSON escaping for newlines ($\backslash$n) and other special characters\\[0.5em]

Base material structure for reference:\\
\hspace*{2em}\{reference\_structures\}\\[0.5em]

Your task:\\
\hspace*{2em}1. Generate \{reproduction\_size\} new structure hypotheses\\
\hspace*{2em}2. Each structure should be stable and physically reasonable\\
\hspace*{2em}3. Format each structure exactly as shown in the input\\[0.5em]

Output your hypotheses below:
}}
\end{center}

\section{Additional Experiments of Crystal Structure Generation and Design}
\label{appendix:multi_opt}
\begin{table*}[h]
\centering
\small
    \begin{tblr}{
    colspec = {ccccccc},
        row{1-3} = {bg=gray!25},
        cell{odd[4-11]}{2-7} = {bg=gray!10}
        }
        \toprule
        \SetCell[r=3]{c}{\textbf{Model}} & 
        \SetCell[r=3]{c}{\textbf{$f$-ele in Parents}} & 
        \SetCell[c=2]{c}{\textbf{Validity}} & & 
        \SetCell[c=3]{c}{\textbf{Metastability}} & & \\
        \cmidrule[lr]{3-4} \cmidrule[lr]{5-7} %
        & & \SetCell[r=2]{c}{Structural} & \SetCell[r=2]{c}{Composition} & M3GNet & \SetCell[c=2]{c}{CHGNet} & \\ %
        \cmidrule[lr]{5} \cmidrule[lr]{6-7} %
        & & & & $E_{\text{d}} < 0.1$ & $E_{\text{d}} < 0.1$ & $E_{\text{d}}<0.03$ \\ %
        \midrule
        CDVAE& ---  & 100.0\% & 86.7\% & 28.8\% & --- & --- \\ %
        \hline[dashed]
        CrystalTextLLM-7B & --- & 96.4\% & 93.3\% & 35.0\% & --- & --- \\ %
        CrystalTextLLM-13B & --- & 95.5\% & 92.4\% & 38.0\% & --- & --- \\ %
        CrystalTextLLM-70B & --- & 99.6\% & 95.4\% & 49.8\% & --- & --- \\ %
        \hline[dashed]
        \SetCell[r=4]{c}{\ours \\ (Llama 3.1-70B)} & Stability & 100.0\% & 79.4\% & 81.1\% & 76.8\% & 56.5\% \\ %
        & Bulk Modulus & 100.0\% & 82.9\% & 27.0\% & 43.3\% & 8.3\%\\ %
        & Multi-turn & 100.0\% & 84.1\% & 70.9\% & 57.1\% & 29.8\% \\ %
        & Weighted Sum & 100.0\% & 88.1\% & 74.0\% & 59.8\%  & 36.5\% \\ %
        & Lexical & 100.0\% & 89.5\% & 84.7\% & 78.0\%  & 59.9\% \\ %
        \bottomrule
        \end{tblr}
\caption{Compare experimental results under various optimization goals. We explored multi-objective optimization for stability and bulk modulus in two different ways. }
\label{tab:multi_obj_exp}
\end{table*}

\label{sec:stable_opt_CSG}

\begin{figure}
    \centering
    \includegraphics[width=\linewidth]{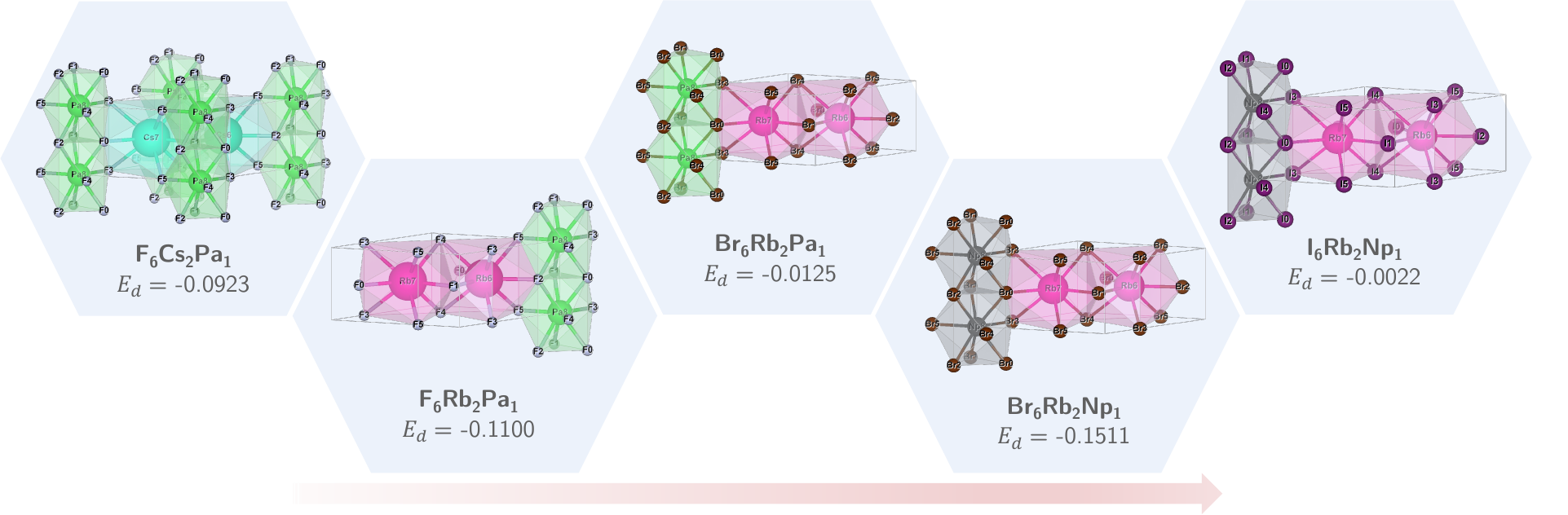}
    \caption{Example trajectory of CSG.}
    \label{fig:csg_trajectory_example}
\end{figure}

\begin{figure}
    \centering
    \begin{minipage}[b]{0.55\textwidth}
    \includegraphics[width=\linewidth]{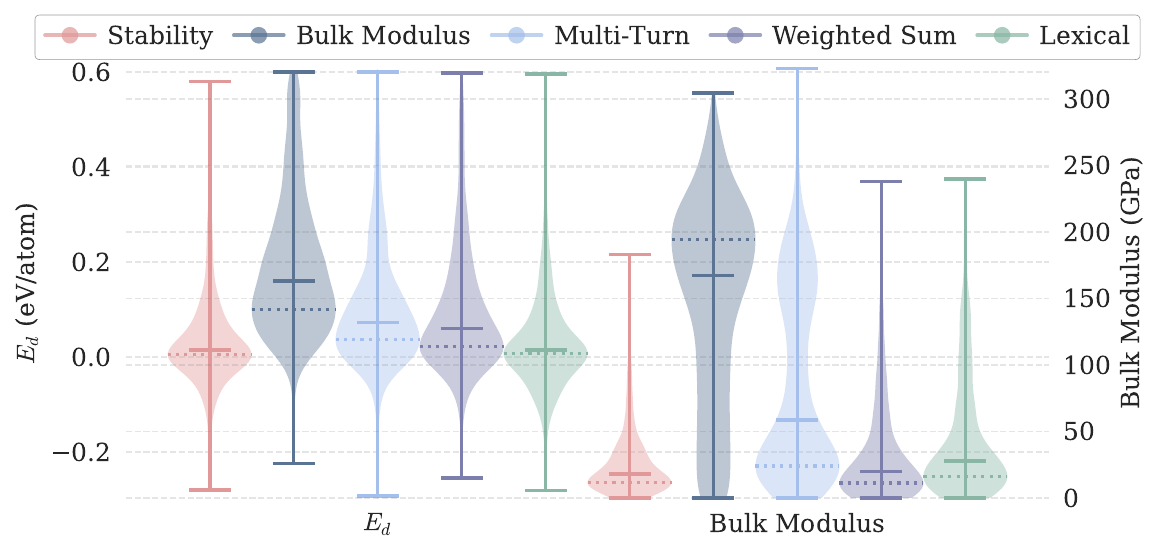}
    \caption{Comparison of optimization strategies targeting different objectives evaluated based on thermodynamic stability (decomposition energy $E_\text{d}$) and mechanical property (bulk modulus).}
    \label{fig:Ed_bulk_comparison_multi_obj}
    \end{minipage}
    \hfill
    \begin{minipage}[b]{0.36\textwidth}
    \includegraphics[width=\linewidth]{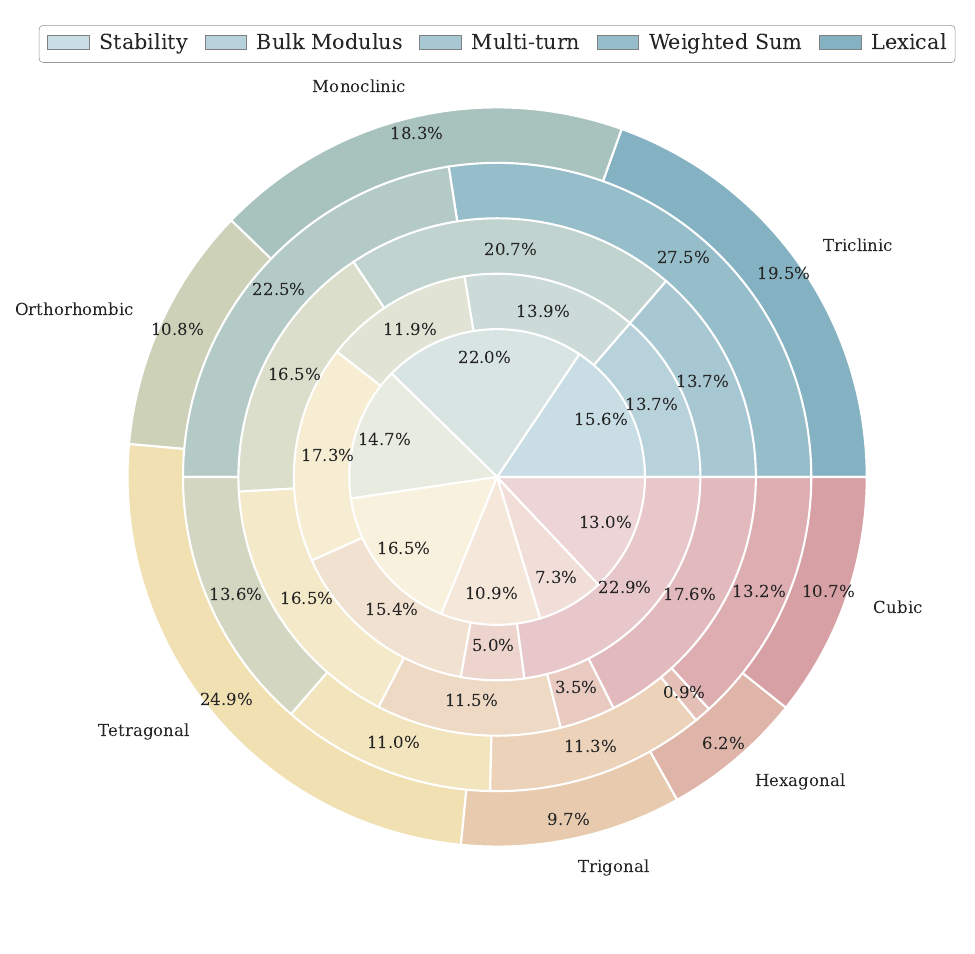}
    \caption{Crystal systems distribution under varied objectives.}
    \label{fig:crystal_systems_comparison_multi_obj}
    \end{minipage}
\end{figure}

The flexibility of our evolutionary pipeline is demonstrated by its ability to guide LLMs in proposing novel crystal structures with diverse mechanical characteristics. We evaluate model performance under five distinct optimization strategies: (1) stability-oriented optimization (``Stability''), (2) property-oriented optimization (``Bulk Modulus''), (3) alternating multi-objective optimization (``Multi-turn''), (4) normalized weighted-sum optimization (``Weighted Sum''), and (5) lexicographic optimization (``Lexical'').
As shown in \cref{tab:multi_obj_exp}, all optimization strategies maintain high metastability rates for the proposed structures, demonstrating that our algorithm can optimize specific properties while maintaining high validity and stability.

\textbf{Stability optimization.}
We visualize an example stability-oriented optimization trajectory in \Cref{fig:csg_trajectory_example}.

\textbf{Bulk modulus optimization.}
To validate the capability of \ours for property-guided generation, we conduct single-property optimization by modifying the selection criteria from decomposition energy ($E_\text{d}$) to bulk modulus. In crystalline solids, bulk modulus serves as a key indicator for designing materials with enhanced mechanical hardness. Our experiments used bulk modulus values derived from the Birch-Murnaghan equation of state as a proof of concept. For more comprehensive materials design applications, this approach can be extended to include elastic tensors from DFT calculations predictions using equivariant graph neural networks \cite{wen2024equivariant}.

\cref{fig:Ed_bulk_comparison_multi_obj} presents the distribution comparison of decomposition energy ($E_\text{d}$) and bulk modulus for structures generated under varied optimization strategies,  revealing distinct performance trade-offs. The bulk modulus optimization generated more structures with larger bulk modulus values, reaching a peak density at 194 GPa compared to only 19 GPa in stability-oriented optimization. However, this enhancement comes at the cost of increased decomposition energy, with the $E_\text{d}$ density peaks shifting from 0.0 eV/atom in stability-oriented optimization to 0.1 eV/atom in bulk modulus optimization, indicating reduced thermodynamically stability across iterations.

\textbf{Multi-objective optimization.}
Beyond single-objective optimization, we explored multi-objective optimization approaches to simultaneously target both thermodynamic stability and mechanical properties using two different multi-objective optimization strategies. 

First we implement an alternating optimization strategy (``Multi-turn''), where the algorithm alternates between optimizing stability and property in successive iterations.
Stability is optimized in the first iteration to set a foundation for property optimization.
For customized multi-objective optimization, the number of iterations for each optimization goal can be adjusted.
As shown in \cref{fig:Ed_bulk_comparison_multi_obj}, this method achieves balanced performance in optimizing stability and bulk modulus, with $E_\text{d}$ centered around 0.037 eV/atom. We observe that bulk modulus distribution separates structures into groups with high mechanical strength at moderate stability versus high stability with lower mechanical strength, suggesting the inherent trade-off in crystal structure generation.

Then we consider a normalized weighted-sum approach that combines both objectives into a single scalar function. We apply min-max normalization to both $E_\text{d}$ and bulk modulus values, then compute the objective as $\mathcal{J} = w_e \cdot \hat{E}_\text{d} + w_b \cdot (1 - \hat{B})$, where $\hat{E}_\text{d}$ and $\hat{B}$ are the normalized values, and $w_e = 0.7$, $w_b = 0.3$ are the weights. This strategy produces crystal structures with bulk modulus centered around 141 GPa and $E_\text{d}$ centered at 0.034 eV/atom.

Then, the ``Lexical'' method prioritizes stability as the primary criterion, only considering bulk modulus for metastable structures ($E_\text{d} < 0.03$ eV/atom). For stable structures, the weighted sum of the two objectives are then being optimized. Low stability structures are penalized to ensure that it remains the dominant factor. While single-objective stability optimization achieves the highest metastability rate of 76.8\%, all multi-objective approaches maintain metastability rates above 50\% while enhancing mechanical properties.

In addition, the analysis of crystal system distributions in \Cref{fig:crystal_systems_comparison_multi_obj} indicates that our framework preserves structural diversity regardless of the optimization objective.

\section{Additional Evaluation on Diversity and Novelty of Generated Structures}
\label{appendix:diversity_novelty}
\begin{figure}
    \centering
    \includegraphics[width=\linewidth]{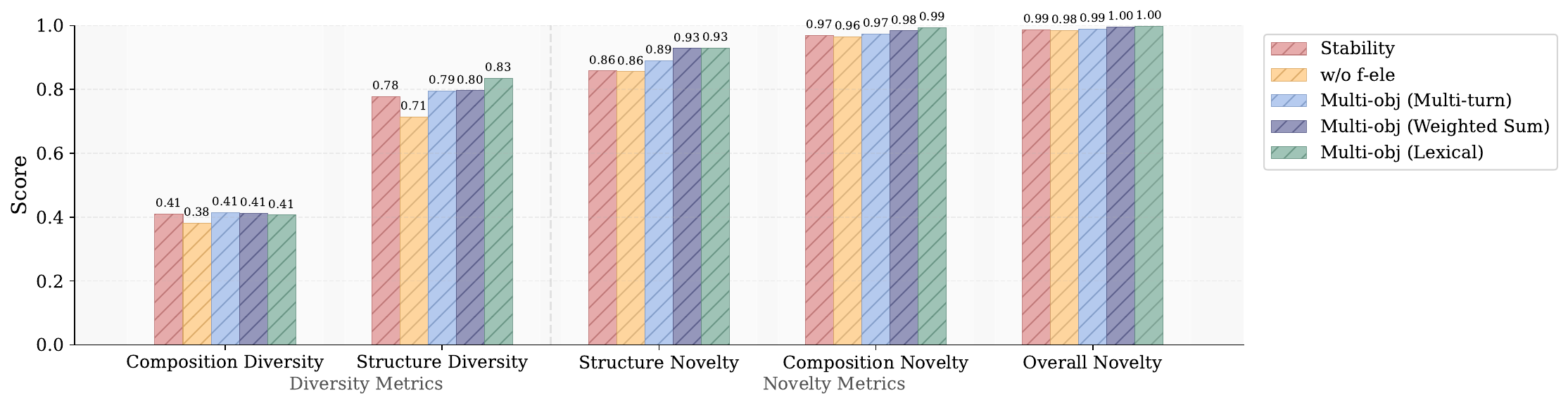}
    \caption{Diversity and novelty evaluation results for structures proposed under different experimental settings.}
    \label{fig:diversity_novelty_comparison}
\end{figure}

We quantitatively evaluate the diversity and novelty of structures generated by our framework across configurations using established metrics from prior work \cite{xie2022crystal, gruver2024llmtime}. Crystal diversity is measured by computing pairwise distances between their structural and compositional fingerprints. Additionally, we apply log normalization to composition diversity for 0-1 scale standardization. The novelty measures the distance between generated samples and their closest neighbors in the extra pool of reference structures. The structural distance cutoff and composition distance cutoff used for novelty calculation are 0.1 and 2 respectively. To align with previous work, all metrics are computed on structures predicted to be metastable. 

The results are summarized in \Cref{fig:diversity_novelty_comparison}. Across different optimization goals, we observe an interesting trade-off between property-specific optimization and novelty, balancing targeted enhancement against chemical space exploration. When optimizing beyond stability alone, such as targeting bulk modulus or performing multi-objective crystal structure design, we observe decreased novelty while diversity remains consistently high across all optimization goals.

While LLMs do tend to favor known stable configurations, our framework's evolutionary approach encourages the exploration of diverse structural motifs, as evidenced by the relatively uniform crystal system distributions.

\paragraph{On S.U.N. metric fairness.}
The S.U.N. (Stable, Unique, Novel) metric~\cite{Zeni2025MatterGen} is convenient but limited: it collapses multi-dimensional quality into counts and depends on the chosen reference/training sets. Our \ours is training-free; therefore, we compute S.U.N. against the MatBench structures used as parents, following DiffCSP/FlowMM protocols. Direct cross-paper comparison is inherently challenging. We therefore position S.U.N. as a supplementary signal alongside multi-MLIP metastability and DFT verification, and we report additional diversity/novelty analyses to capture broader scientific value.

To further assess structural diversity, we extended our evaluation with additional metrics, including the S.U.N. Rate (Stable, Unique, Novel) aligning to MatterGen and diversity and novelty regarding composition and structure in \Cref{tab:sun_rate}. High diversity and novelty are consistently achieved by our approach under various settings.

We provide comprehensive evaluation across all S.U.N. dimensions: For Stability, we report metastability rates computed with multiple MLIPs (CHGNet and M3GNet for CSG and additional Orb-v3 for CSP) and DFT-calculated stability rates, providing more rigorous thermodynamic assessment than S.U.N.'s single stability threshold. For Uniqueness, we provide detailed analysis including space group distributions and crystal system diversity, offering deeper insights into the true diversity of generated structures. For Novelty, we measure compositional and structural novelties and demonstrate elemental co-occurrence pattern shifts that indicate genuine exploration of novel composition space regions. 

\noindent\textit{Methodological note.} In \Cref{tab:sun_rate}, the ``Stable'' criterion follows FlowMM: we first relax with CHGNet and then verify stability with DFT at the 0.0 eV/atom threshold on the convex hull. This aligns the protocol for fair comparison while acknowledging that our training-free setup necessitates computing uniqueness/novelty against MatBench parents.

\section{Time Overhead}
\label{appendix:time_overhead}

\begin{wraptable}{r}{0.5\textwidth}
    \centering
    \small
    \begin{tblr}{
        colspec = {l X[c,m] X[c,m]},
        row{1} = {bg=gray!25},
        cell{odd[2-3]}{1-3} = {bg=gray!10},
        width = 0.48\textwidth
    }
        \toprule
        \textbf{Model} & \textbf{Avg. Generation Time (s)} & \textbf{Avg. Evaluation Time (s)} \\
        \midrule
        70B & 55.99 & 6.36 \\
        8B & 53.40 & 10.80 \\
        \bottomrule
    \end{tblr}
    \caption{Time overhead for different model sizes}
    \label{tab:time_overhead}
\end{wraptable}

The efficiency of our method is primarily determined by the hardware resources and model size used. As reported in Table~\ref{tab:time_overhead}, with Llama-3.1-70B-Instruct running on 4 A6000 GPUs and a population size of 100, the average time to propose one valid unique structure is 62.35 seconds. In comparison, CrystalTextLLM takes 51.6 seconds on average to propose one valid structure. Our time overhead can be further optimized with improved computational resources.

\noindent\textit{Inference cost rationale.} We report average time per successfully generated valid structure, rather than raw FLOPs, because larger models achieve higher success rates per attempt. Although a 70B model has higher per-step compute than an 8B model, its greater validity rate often yields lower cost per successful sample.

\section{Additional Discussion on Ablation Analysis. }
\label{appendix:ablation}
\subsection{Base LLM Analysis}

\textbf{Model Scale Effects.}
We evaluate the framework across different LLM configurations, including pre-trained and fine-tuned Llama 3.1 models (8B and 70B). \Cref{tab:finetuned_exp} demonstrates that model capability significantly impacts CSG metastability. The 70B model achieves 76.8\% metastability in our full framework compared to 27.7\% for the 8B model, indicating that crystallographic knowledge emerges at sufficient model scale.

\begin{table}[ht]
\centering
\renewcommand{\arraystretch}{0.9} 
\SetTblrInner{rowsep=0pt} 
\setlength{\aboverulesep}{0.1ex}
\setlength{\belowrulesep}{0.1ex}
\begin{tblr}{
colspec = {ccccc},
    row{1-2} = {bg=gray!25},
    cell{even[3-14]}{3-5} = {bg=gray!10},
    cells= {font=\small},
    }
    \toprule
    \SetCell[r=2]{c}{\textbf{Base Model}} & 
    \SetCell[r=2]{c}{\textbf{Fine-Tuning?}} & 
    \SetCell[r=2]{c}{\textbf{Prompting}\\ \textbf{Strategy}} & 
    \SetCell[c=2]{c}{\textbf{Metastable(\%)} (CHGNet)} &\\
    \cmidrule[lr]{4-5}
    & & & $E_{\text{d}} < 0.1$ eV/atom & $E_{\text{d}}<0.03$ eV/atom \\
    \midrule
    \SetCell[r=4]{c}{{Llama 3.1 8B}} & \SetCell[r=2]{c}{\xmark}  & Zero-shot  & 0.0 & 0.0 \\
    & & \ours  & 27.7 & 10.0 \\
    \hline[dashed]{2-5}
    & \SetCell[r=2]{c}{\cmark (8 bit)}  & Zero-shot  & 0.0 & 0.0 \\
    & & \ours  & 45.5 & 22.7 \\
    \midrule
    \SetCell[r=4]{c}{{Llama 3.1 70B}} & \SetCell[r=2]{c}{\xmark}  & Zero-shot  & 25.8 & 12.9 \\
    & & \ours  & 76.8 & 56.5 \\
    \hline[dashed]{2-5}
    & \SetCell[r=2]{c}{\cmark (4 bit)}  & Zero-shot  & 13.9 & 2.8 \\
    & & \ours  & 66.0 & 48.0 \\
     \bottomrule
\end{tblr}
\caption{Meta-stability comparison of prompting strategy across models with and without fine-tuning.}
\label{tab:finetuned_exp}
\end{table}

\textbf{Fine-tuning Integration.}
Fine-tuned models show substantial improvements when integrated with our evolutionary framework. The 8B fine-tuned model achieves 45.5\% metastability (from 27.7\% baseline), while the 4-bit quantized 70B model maintains 66.0\% metastability despite compression constraints. See \Cref{tab:finetuned_exp} for detailed results and prompting strategies. Importantly, our information value metric demonstrates that both fine-tuned and pre-trained models integrate seamlessly into the evolutionary framework, with performance scaling according to base model capability.

\subsection{Generation Strategy Comparison}
We extend the comparison to two generation strategies: zero-shot (w/o reference structures nor evolution) and our evolutionary framework. Zero-shot approaches fail for 8B models and achieve only 25.8\% metastability for 70B models. Our evolutionary framework overcomes these limitations by systematically exploring chemical space while maintaining high stability rates.

The comprehensive analysis reveals several critical insights: (1) Reference structures simultaneously accelerate convergence and promote structural diversity exploration across multiple dimensions, (2) Evolutionary iterations are essential for practical generation volumes and sustained quality optimization, (3) The synergistic combination of both components achieves optimal performance that exceeds individual contributions, (4) Our information value metric successfully captures complex trade-offs that single-dimensional metrics miss, enabling objective comparison across methodologies. These findings establish a robust evaluation framework for crystal structure generation while demonstrating the effectiveness of our evolutionary approach.

\section{Impact of Structure Relaxation}
\label{appendix:relaxation}
\begin{figure}
    \centering
    \includegraphics[width=\linewidth]{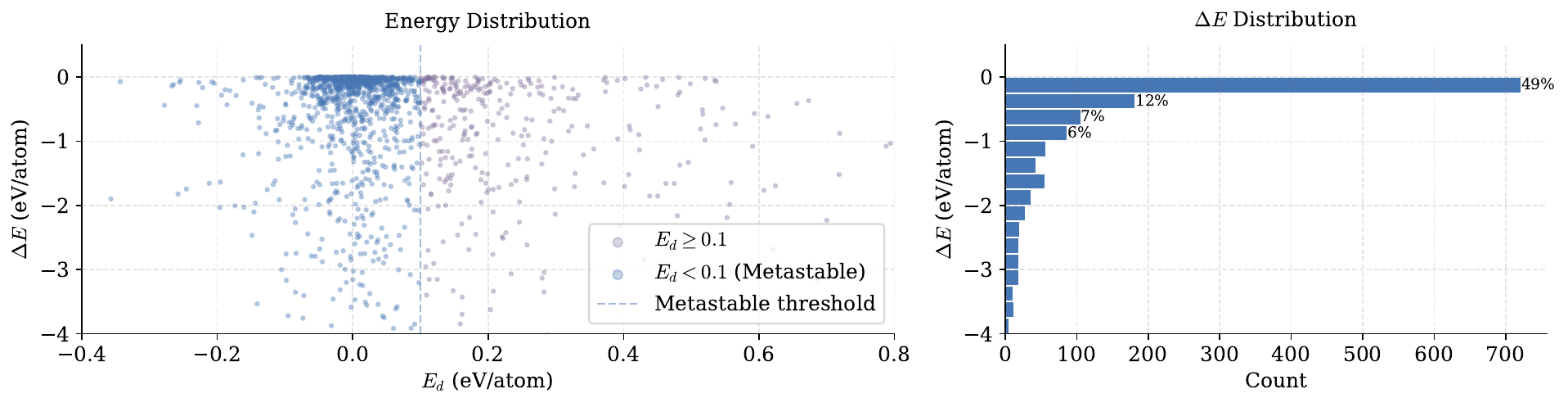}
    \caption{Distribution of energy change $\Delta E$ before/after structural relaxation and decomposition energy ($E_\text{d}$) for structures proposed by LLM, evaluated using the pretrained CHGNet.}
    \label{fig:delta_e_e_d_distribution}
\end{figure}
To measure the contribution of structural relaxation in our framework, we introduce a quantity $\Delta E$ to represent the energy difference after and before structural relaxation using CHGNet. \Cref{fig:delta_e_e_d_distribution} reveals that the majority of the proposed structures proposed by LLMs exhibit a relatively small $\Delta E$, with 61.2\% showing minimal energy changes ($|\Delta E| < 0.5$ eV/atom) during relaxation. This distribution indicates that our framework generates physically meaningful structures that are already close to their local energy minima, requiring only modest refinements through relaxation.

\section{Impact of Structural Representation}
\label{appendix:format}
\begin{table*}
\centering
\small
\begin{tblr}{
colspec = {ccccc},
    row{1} = {bg=gray!25},
    cell{odd[2-8]}{1-5} = {bg=gray!10}
    }
    \toprule
    \textbf{Method} & \textbf{Primary Format}  & \textbf{Generative} & \textbf{Model} & \textbf{Training}  \\
    \midrule
    CDVAE \cite{xie2022crystal} & 3D & Diffusion & GNN & Training  \\
    MatterGen \cite{Zeni2025MatterGen} & 3D & Diffusion &  GNN & Training \\
    \citet{flam2023language} & 3D & AR & Transformer & Training \\
    DiffCSP \cite{jiao2024crystal} & 3D & Diffusion & GNN & Training \\
    CrystalTextLLM \cite{gruver2024llmtime} & Text/CIF & LLM & Transformer & Fine-tuning  \\
    FlowMM \cite{sriram2024flowllm} & 3D & Flow & GNN & Training  \\
    \midrule
    \ours (Ours) & Text/CIF/POSCAR & LLM & Llama 3.1 & N/A \\
    \bottomrule
\end{tblr}
\caption{A collection of generative models on computational materials discovery. Training denotes if training/fine-tuning is required on materials databases. }
\label{tab:baseline_comparison}
\end{table*}

A number of computational methods has emerged for crystal structure generation using machine learning approaches, as shown in \Cref{tab:baseline_comparison}. Most methods represent crystal structures using 3D information processed through either Graph Neural Networks (GNN) or Transformer architectures, employing various generative strategies like diffusion models or autoregressive approaches. More recently, text-based formats and Large Language Models (LLMs) have emerged as an alternative approach, signaling a promising shift in crystal structure generation and analysis techniques.

\begin{figure}[ht]
    \centering
    \begin{minipage}[b]{0.48\linewidth}
        \centering
        \includegraphics[width=\linewidth]{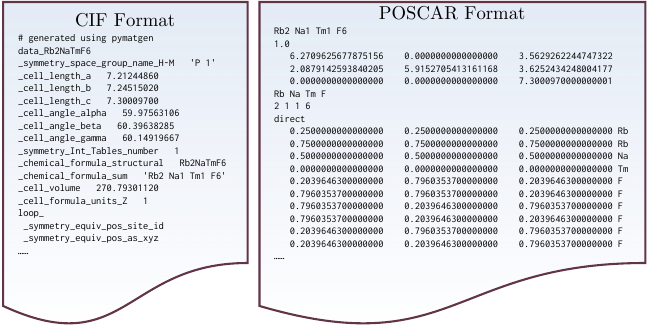}
        \caption{Structure string examples of CIF format and POSCAR format.}
        \label{fig:format_example}
    \end{minipage}
    \hfill
    \begin{minipage}[b]{0.48\linewidth}
        \centering
        \includegraphics[width=\linewidth]{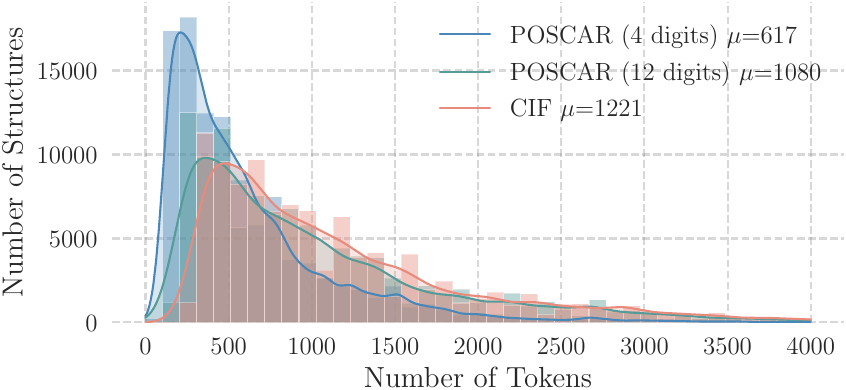}
        \caption{Token efficiency comparison under CIF formatting and POSCAR formatting for the precision of 4 and 12 decimal. $\mu$ indicate the mean of token lengths.}
        \label{fig:format_token_comparison}
    \end{minipage}
\end{figure}

The encoding of crystallographic structures into text-based format is essential for LLM processing. We investigated the impact of different structural representation strategies on generation efficiency and performance: CIF format and POSCAR format with either 4 or 12 decimal places of precision. See \Cref{fig:format_example} for examples.

First, we examine the token efficiency by analyzing the MatBench dataset for token length distribution as shown in \Cref{fig:format_token_comparison}. The distribution indicates that the POSCAR format with 4 decimal places offers the most token-efficient representation while maintaining reasonable precision, followed by the POSCAR with 12 digits and CIF format. CIF format requires more tokens than POSCAR format, given that CIF uses a more verbose structure and additional metadata. 

\begin{table}[t]
\centering
\small
\SetTblrInner{rowsep=0.95pt}
\begin{tblr}{
colspec = {cccc},
    row{1} = {bg=gray!25},
    cell{odd[2-5]}{1-4} = {bg=gray!10}
    }
    \toprule
    Format & \# Unique / \# Total generated & $E_{\text{d}} < 0.1 $ eV/atom & $E_{\text{d}}<0.03$ eV/atom   \\
    \midrule
    POSCAR (4) & 76.7\% & 75.4\%  & 55.3\%  \\
    POSCAR (12) & 72.3\% & 76.8\% & 56.5\%  \\
    CIF  & 75.1\% & 68.9\%  & 49.5\%  \\
    \bottomrule
\end{tblr}
\caption{Proportion of unique structures and their CHGNet-predicted metastability using different structure formats.}
\label{tab:format_exp}
\end{table}

Performance evaluation shown in \Cref{tab:format_exp} suggests that POSCAR formatting in 12 decimal places demonstrates slightly better overall performance in the rate of (meta)stability of generated structures under different criteria ($E_\text{d} < 0.03$ or $0.1$ eV/atom). 
Therefore, we employ POSCAR of 12 decimal places as a trade-off results of token efficiency and informativeness. The marginal difference across format may be attributed to the crystallographic data exposed to the LLMs during pre-training.
However, it is noteworthy that performance differences across formats remain modest, suggesting the resilience of our approach across different structural representations. 

\section{Hyper-Parameter Studies}
\label{appendix:hyperparameters}
\begin{table}[ht]
\centering
\small
\begin{tblr}{
colspec = {cccc},
    row{1} = {bg=gray!25},
    cell{odd[2-6]}{1-4} = {bg=gray!10}
    }
    \toprule
    Reproduction Configuration & \# Unique / \# Total generated & $E_{\text{d}} < 0.1 $eV/atom & $E_{\text{d}}<0.03$ eV/atom  \\
    \midrule
    $1 \rightarrow 5$ & 56.5\% & 79.8\% & 56.4\% \\
    $2 \rightarrow 5$ & 72.3\% & 76.8\% & 56.5\% \\
    $2 \rightarrow 2$ & 86.3\% & 74.8\% & 54.3\%\\
    $5 \rightarrow 5$ & 92.7\% & 72.3\% & 47.3\%\\
    $5 \rightarrow 2$ & 95.5\% & 68.3\% & 46.1\%\\
    \bottomrule
    \end{tblr}
    \caption{Proportion of unique structures and their CHGNet-predicted metastability under varying reproduction configurations.}
\label{tab:hyperparameter_exp}
\end{table}

\begin{table}[ht]
\centering
\small
\begin{tblr}{
colspec = {cccc},
    row{1} = {bg=gray!25},
    cell{odd[2-5]}{1-4} = {bg=gray!10}
    }
    \toprule
    LLM Temperature & \# Unique / \# Total generated & $E_{\text{d}} < 0.1$ eV/atom & $E_{\text{d}}<0.03$ eV/atom   \\
    \midrule
    0.95 & 72.3\% & 76.8\% & 56.5\% \\
    0.7 & 70.7\% & 75.4\% & 56.6\%\\
    0.5 & 70.7\% & 71.2\% & 51.4\%\\
    0.2 & 69.8\% & 70.3\% & 50.2\%\\
    \bottomrule
    \end{tblr}
    \caption{Proportion of unique structures and their CHGNet-predicted metastability with different LLM temperatures.}
    \label{tab:temperature_exp}
\end{table}

\textbf{Reproduction parameters.}
Our training-free evolutionary framework significantly reduces hyperparameter sensitivity compared to traditional machine learning methods.
The reproduction phase introduces several key hyper-parameters that influence LLMs' generation behavior and efficiency, including population size ($K$), context size ($C$), and children size ($c$). Our baseline configuration ($C=2$, $c=5$) leverages the Llama 3.1 (70B) model to achieve balanced performance, generating 72.29\% unique structures while maintaining high stability rates.

Analysis of parent-to-children ratios reveals that increasing parent diversity ($C=5$, $c=2$) can enhance composition uniqueness of generated structures to 95.49\%, though at the price of slight decrease in stability, as presented in \Cref{tab:hyperparameter_exp}. Conversely, results with single parent demonstrates that crossover between multiple parent structures is beneficial for maintaining structural diversity and stability in the generation process. Overall, we believe that higher parent-to-children ratios can lead to better overall quality in generated structures.

Our analysis also reveals that larger population sizes $K$ can maintain high stability and validity rates comparable to smaller populations. One potential benefit of increasing population size is the diversity introduced in the iteration process, which can alleviate the overpopulation of $f$-ele structures but also lead to higher compositional diversity. However, the increased diversity is offset by higher rates of structural duplication across iterations, suggesting earlier convergence may be needed. Our findings above enable application-specific optimization of the framework's parameters.

\textbf{Model temperature.}
The temperature hyper-parameter controls sampling randomness in language models by scaling the logits before softmax transformation. Higher temperatures flatten the probability distribution, increasing sampling diversity, while lower temperatures concentrate probability mass on the most likely tokens.  While temperature is commonly associated with model creativity, with higher temperatures generally producing slightly more novel outputs~\cite{Peeperkorn2024IsTT}, this relationship remains an active area of research.

Crystal structure generation is a creative task that requires exploring diverse structural possibilities while maintaining physical validity. We employed an LLM inference temperature of 0.95 in our baseline experiments to facilitate broader structural exploration while maintaining reasonable generation stability. 
In \Cref{tab:temperature_exp}, we present the metastability evaluated by CHGNet for structures generated with different LLM temperatures. At the temperature of 0.95, the LLM generated 76.81\% metastable structures with $E_\text{d} < 0.1$ eV/atom as evaluated by CHGNet. Reducing the temperature to 0.7 maintained robust performance, producing 75.38\% metastable structures. Further lowering the temperature to 0.5 yields 71.18\% metastable structures. If we choose $E_\text{d} < 0.03$ eV/atom as the stability criterion, the percentage of qualifying structures at temperatures 0.95, 0.7, 0.5 and 0.2 are 56.5\%, 56.6\%, 51.4\% and 50.2\%  respectively. 
 The consistent high stability rates across temperature settings demonstrate the robustness of our pipeline to LLM hyper-parameter variations.

\section{Crystal Structure Prediction Examples.}
\label{appendix:CSP}
\begin{figure}[htb]
    \centering
    \includegraphics[width=\linewidth]{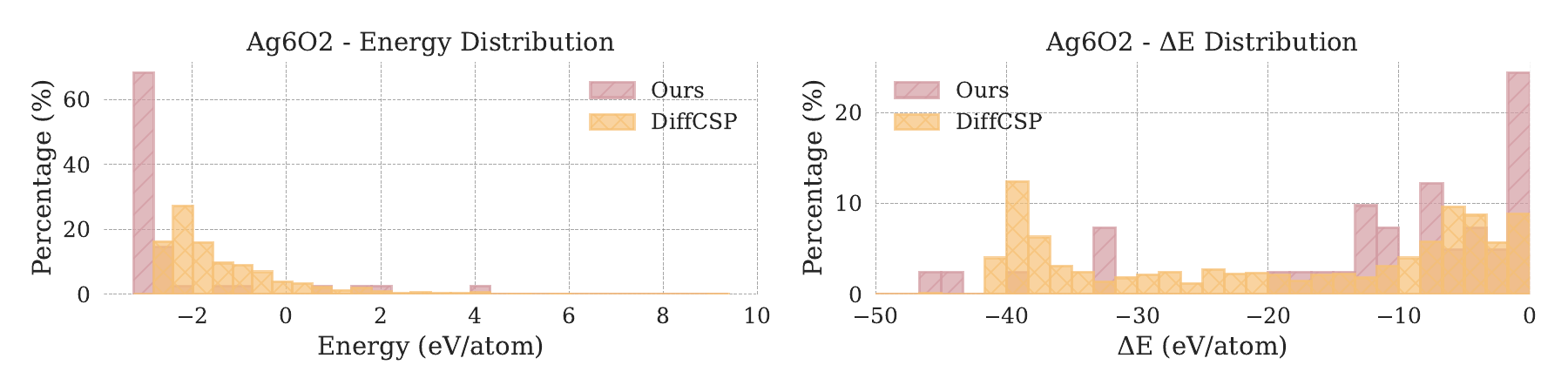}
    \includegraphics[width=\linewidth]{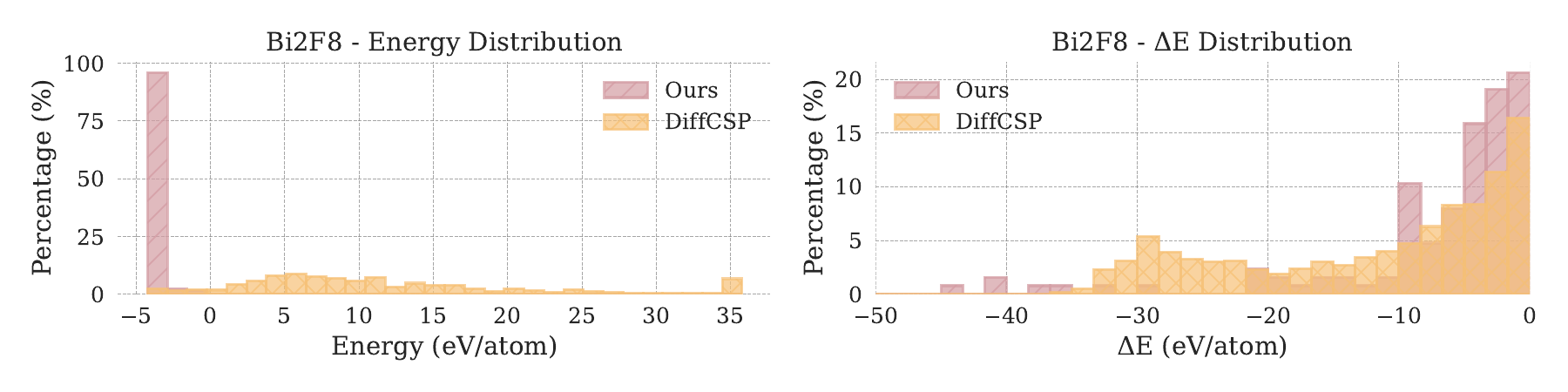}
    \caption{Energy and $\Delta E$ distribution of CSP results for \ce{Ag6O2} and \ce{Bi2F8}.}
    \label{fig:CSP_energy_deltae}
\end{figure}

\begin{figure}[htb]
    \centering
    \includegraphics[width=\linewidth]{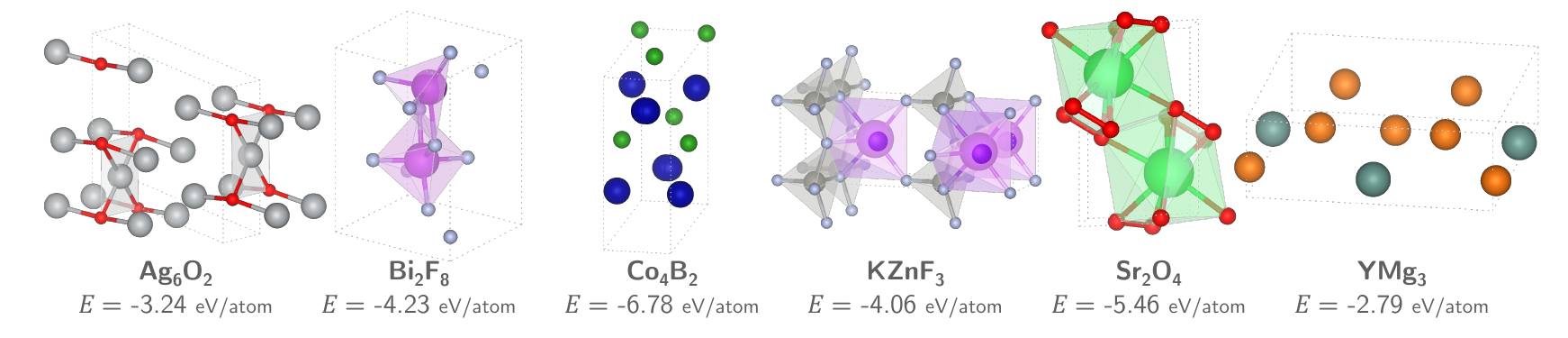}
    \caption{CSP examples.}
    \label{fig:CSP_examples}
\end{figure}

We further evaluate \ours on crystal structure prediction across multiple compositions. Crystal structure prediction fundamentally involves searching for the optimal atomic arrangement that minimizes the system's energy for a given composition.

For CSP task, we first apply compositional constraints to filter seed structures, and then execute \ours for 10 iterations. This process enables the LLM to predict crystal structures by referencing from optimized structures with similar compositions.

To benchmark the effectiveness of our framework, we compare our results against structures generated and optimized by DiffCSP. We first used DiffCSP to sample 100 candidate structures for each composition. We then trained an energy predictor model on the MP-20 dataset for 1000 epochs to learn formation energy predictions. Then we applied the energy-guided optimization procedure to refine the structures with the trained energy model, generating 10 optimized variants for each initial structure. Orb-v3 \cite{rhodes2025orbv3} are employed to evaluate the energy of the relaxed structure from both sources.

\ours successfully predicts structures for various compositions, including \ce{Ag6O2}, \ce{Bi2F8}, \ce{Co4B2}, \ce{KZnF3}, \ce{Sr2O4}, and \ce{YMg3}. \Cref{fig:CSP_examples} presents representative examples of these predicted crystal structures, which achieve lower energies than the best structures predicted by DiffCSP.

To further evaluate the two approaches, we analyzed the energy and $\Delta E$ distributions of the generated structures. As shown in \Cref{fig:CSP_energy_deltae}, a large portion of structures predicted by DiffCSP experience substantial structural changes during relaxation, as indicated by large $|\Delta E|$ values. In contrast, our LLM-generated structures demonstrate superior initial stability, requiring minimal relaxation. This suggests that our initial configurations are already close to local energy minima.

\section{Algorithm Details}
\label{appendix:algorithm}

\begin{algorithm}
\small
\caption{The \ours Framework}
\label{algo:algorithm}
\begin{algorithmic}[1]
\Require Population size $K$, parent size $P$, reproduction size $C$, number of iterations $N$, known stable structures $\mathcal{D}$, oracle function $O$, extra pool $\mathcal{R}$
\LComment{Initialization}
\State Form population $\mathcal{P}_0$ by sampling $K$ groups of $P$ structures from $\mathcal{D}$
\State Initialize structure collection $\mathcal{S} \gets \varnothing$
\For{$i \gets 0, 1, \cdots, (N-1)$}
    \LComment{LLM-guided reproduction}
    \State Generate prompts from parent structures in $\mathcal{P}_i$
    \State Obtain offspring structures $\mathcal{C}_i$ via LLM inference and parsing
    
    \LComment{Structure evaluation}
    \State Relax structures $\mathcal{C}_i \gets \text{CHGNetRelax}(\mathcal{C}_i)$
    \State Calculate decomposition energy $E_\text{d}$ and properties
    \State Evaluate objective scores using oracle function $O$
    \State Update structure collection $\mathcal{S} \gets \mathcal{S} \cup \mathcal{C}_i$
    
    \LComment{Selection}
    \State Form candidate pool from parents $\mathcal{P}_i$, offspring $\mathcal{C}_i$, and extra pool $\mathcal{R}$
    \State Select top-$(K\times P)$ structures based on objective scores from the candidate pool
    \State Construct next parent groups $\mathcal{P}_{i+1}$
\EndFor
\State Validate final structures via DFT
\State \Return cumulated structures $\mathcal{S}$
\end{algorithmic}
\end{algorithm}

\section{Details of Machine Learning Interatomic Potentials}
\label{appendix:MLIP}

A significant breakthrough in addressing computational cost challenges has emerged through the development of machine learning interatomic potentials (MLIPs) trained based on high-fidelity quantum mechanical calculations (e.g., DFT)~\cite{zhang2021phase, batzner2022nequip, lopez2023aenet, equiformer_v2, cheng2024cartesian, du2023m, du2024new,yin2025alphanetscalinglocalframebased}. In MLIPs, the total energy is expressed as a sum of atomic contributions, where each atom's energy depends on its local environment including the atomic coordinates and chemical species of neighboring atoms within a cutoff radius:
\begin{equation}
\hat{E} = \sum_i^n \phi(\{\vec{r}_j\}_i, \{C_j\}_i), \quad \hat{\boldsymbol{f}}_i = - \frac{\partial \hat{E}}{\partial \boldsymbol{r}_i}, \quad \boldsymbol{\sigma} = \frac{1}{V}\frac{\partial \hat{E}}{\partial \boldsymbol{\varepsilon}}.
\end{equation}
Here, $\phi$ is a learnable function that maps the set of position vectors $\{\vec{r}_j\}_i$ and chemical species $\{C_j\}_i$ of the neighboring atoms $j$ to the energy contribution of atom $i$. The forces $\boldsymbol{f}_i$ and stress $\boldsymbol{\sigma}$ are calculated via auto-differentiation of the total energy with respect to the atomic Cartesian coordinates and strain. Recent advances have demonstrated that MLIPs, trained on extensive density functional theory (DFT) calculations accumulated over the past decade across diverse materials systems, exhibit remarkable transferability in performing atomistic simulations across various material and chemical systems. These broadly applicable potentials are known as universal MLIPs (uMLIPs)~\cite{chen2022m3gnet, deng2023chgnet, batatia2023foundation, park_scalable_2024}. By leveraging uMLIPs as surrogate energy models, researchers can rapidly optimize crystal structures and obtain structure-energy relationships for assessing thermodynamic stability. By leveraging uMLIPs as surrogate energy models, one can rapidly optimize crystal structure and obtain the structure-energy relationships for assessing thermodynamic stability. Recent benchmark studies, including MACE~\cite{batatia2023foundation}, DPA-1~\cite{zhang2024dpa1} and JMP (joint multi-domain pretraining)~\cite{shoghi2023molecules}, have demonstrated the high accuracy of these uMLIPs in predicting crystal thermodynamical stability, particularly for industrial-scale implementations trained on millions of compounds and non-equilibrium atomic configurations \cite{merchant2023GNoME, barroso2024open, yang2024mattersim}. 

To accelerate the oracle function evaluation in the evolutionary iterations, we performed all structure relaxations with the \texttt{FIRE} optimizer \cite{Bitzek2006_FIRE} over the potential energy surface provided by CHGNet, where the atom positions, cell shape, and cell volume were optimized to reach converged interatomic forces of 0.1 eV/atom \cite{deng2023chgnet}. The output energy prediction is directly compatible with the Materials Project phase diagrams with the \texttt{MaterialsProject2020Compatibility}~\cite{wang2021dft}.

\end{document}